# Dynamic Redundancy-aware Blockchain-based Partial Computation Offloading for the Metaverse in In-network Computing

Ibrahim Aliyu, *Member, IEEE*, Cho-Rong Yu, Tai-Won Um and Jinsul Kim, *Member, IEEE*

*Abstract*—The computing in the network (COIN) paradigm has emerged as a potential solution for computation-intensive applications like the metaverse by utilizing unused network resources. The blockchain (BC) guarantees task-offloading privacy, but cost reduction, queueing delays, and redundancy elimination remain open problems. This paper presents a redundancy-aware BC-based approach for the metaverse's partial computation offloading (PCO). Specifically, we formulate a joint BC redundancy factor (BRF) and PCO problem to minimize computation costs, maximize incentives, and meet delay and BC offloading constraints. We proved this problem is NP-hard and transformed it into two subproblems based on their temporal correlation: real-time PCO and Markov decision process-based BRF. We formulated the PCO problem as a multiuser game, proposed a decentralized algorithm for Nash equilibrium under any BC redundancy state, and designed a double deep Q-network-based algorithm for the optimal BRF policy. The BRF strategy is updated periodically based on user computation demand and network status to assist the PCO algorithm. The experimental results suggest that the proposed approach outperforms existing schemes, resulting in a remarkable 47% reduction in cost overhead, delivering approximately 64% higher rewards, and achieving convergence in just a few training episodes.

## I. INTRODUCTION

T HE metaverse is envisaged to provide an immersive experience but faces constraints with user equipment (UE) and mobile edge computing (MEC) limitations [1, 2]. Optimizing resource allocation is vital for metaverse performance while guaranteeing user security. Conversely, the COIN paradigm minimizes delays and optimizes user experience using underutilized network resources [3, 4]. However, enabling COIN increases power consumption, driving a need for optimizing time delay and energy use in metaverse tasks

The blockchain (BC) provides decentralized data privacy and secured communication but its full data redundancy creates scalability issues, such as high storage demands, impacting decentralization For example, Bitcoin's BC exceeds 477 GB, requiring 10 000 nodes to have about 4.6 PG storage to participate in the network [5]. Addressing data redundancy is crucial for efficient computational offloading, involving complex optimization considering BC costs, mining incentives, decentralization, and offloading expenses. Thus, integrating metaverse with other promising technologies, such as COIN and BC presents numerous opportunities and research challenges to support next-generation services.

### A. Related Work

This section presents related work on task offloading and trust problems in COIN.

*Metaverse Task Offloading.* Previous studies have extensively explored task offloading, especially in MEC networks [6-8].For example, Jošilo and Dán [9] addressed dynamic slice allocation for latency-sensitive task offloading in edge computing. However, most of the existing literature has primarily concentrated on indivisible tasks, often overlooking the significance of divisible tasks, which are pivotal in the metaverse scenario [8, 10, 11]. Subtask execution is vital in the metaverse due to latency and data intensity. While Alriksson et al. [12] proposed XR task decoupling within 5G networks, optimizing task offloading modes for large-scale metaverse deployments remains an ongoing challenge. Similarly, Tütüncüoğlu, et al. [13] discussed subtask offloading in serverless edge computing, and Zhang, et al. [14] addressed subtask dependency offloading in MEC. Nonetheless, these studies predominantly revolved around binary offloading decisions and static user-resource topologies, often overlooking concerns related to secure offloading.

*Blockchain-based Offloading Schemes*. Task offloading typically encompasses full or partial offloading. Blockchain (BC) has been introduced to ensure secure task offloading [15].

This research was supported by the Culture, Sports and Tourism R&D Program through the Korea Creative Content Agency grant funded by the Ministry of Culture, Sports and Tourism in 2022 (Project Name: Development of real-time interactive metaverse performance experience platform technology on the scale of a large concert hall Project Number: RS-2022-050002, Contribution Rate: 100%); and partly supported by the Institute of Information and Communications Technology Planning and Evaluation (IITP) grant funded by the Korean government (MSIT) (No.2021-0-02068, Artificial Intelligence Innovation Hub).

Corresponding authors: Tai-Won Um (email: stwum@jnu.ac.kr) and *Jinsul Kim* (email: jsworld@jnu.ac.kr).

Ibrahim Aliyu, and Jinsul Kim are with the Department of ICT Convergence System Engineering, Chonnam National University, Gwangju 61186, Korea.

Cho-Rong Yu is with : Hyper-Reality Metaverse Research Laboratory, Content Research Division, Electronics and Telecommunications Research Institute, 218, Gajeong-ro, Yuseong-gu, Daejeon, 34129, Korea.

Tai-Won Um is with the Graduate School of Data Science, Chonnam National University, Gwangju 61186, Korea.

For instance, Lang, et al. [16] delved into cooperative task offloading within vehicular edge computing through BC and proposed a consensus mechanism to guarantee secure handover. Other studies explore schemes encompassing trust-based collaboration among edge nodes, Hyperledger Fabric frameworks for secure task offloading in MEC, and incentive mechanisms for mining task offloading in fog or cloud computing [17-20]. Despite these notable efforts, the majority of BC-based systems have primarily centered around atomic (indivisible) tasks and relied on full BC redundancy.

***Data Redundancy in the Blockchain.*** The scalability issues of BC stem from its inherent append-only data structure, which necessitates increasingly larger storage resources [21]. The practice of maintaining full BC nodes locally results in excessive data redundancy, compelling many nodes to transition to light nodes. This transition poses a significant threat to decentralization and hampers the practical application of BC in offloading scenarios. In addition, various strategies have been devised to mitigate the scalability problems of BC, encompassing multiple node cooperative storage (MNCS), data reduction, and sharding [21]. MNCS schemes, such as the full and lightweight node approach, aim to alleviate data redundancy [22, 23]. Data reduction techniques encompass data compression and pruning [24-26], while sharding partition nodes enhances throughput and alleviates storage pressure [21]. However, Optimizing the replica copy of a transaction in a BC network for multiuser scenarios is an enormous problem to overcome, given factors like cross-shared communication overhead, shard management, BC costs, and overall offloading expenses.

This section outlines previous research on metaverse task offloading, blockchain-based offloading schemes and data redundancy. It highlights their limitations concerning divisible tasks and data redundancy in BC. Table I provides a summarized overview of related work.

### B. Motivation and Contributions

As evident from the related work section, substantial efforts have been dedicated to enhancing task offloading and resource allocation, with a focus on the efficacy of Mobile Edge Computing (MEC) and fog computing for various applications [6-8]. Nonetheless, the rise of computation-intensive applications like the metaverse raises questions about the internet's capacity to handle computation efficiently. In the Metaverse, offloading is essential for tasks like AR/VR rendering on devices, enabling AI for non-player characters (NPCs), and managing data with many subscribers. The COIN paradigm seeks to accelerate computation. Although COIN is in its early stages, integrating trust and security into its offloading poses a valuable challenge. BC is well-known for its trust and security through its distributed ledger, but it grapples with data redundancy as nodes replicate offloaded data. The BC can facilitate resource and service trading between user terminals, servers, COIN/Fog/Edge nodes, and item/information exchange among metaverse avatars in the metaverse. Previous research has explored BC for atomic tasks and edge computing. In the metaverse, frequent partial offloading of millions of AI-based NPCs avatars, for instance,

| Related work | Method | Offloading mode | BC Redundancy | BC mining incentive | Metaverse | COIN |
|---|---|---|---|---|---|---|
| 2018 [27] | TS | NA | ✓ | × | × | × |
| 2018 [28] | TS | NA | ✓ | × | × | × |
| 2018 [29] | Game theory | NA | × | ✓ | × | × |
| 2020 [17] | Lyapunov optimization | Partial | × | × | × | × |
| 2020 [30] | Online-learning | Full | × | ✓ | × | × |
| 2021 [18] | | | | | | |
| 2021 [31] | Game theory | Full | × | ✓ | × | × |
| 2021 [20] | Hyperledger fabric | Full | × | × | × | × |
| 2021 [32] | Berger model, service justice | Full | × | ✓ | × | × |
| 2022 [15] | RL | Full | × | ✓ | × | × |
| 2023 [16] | RL | Full | × | ✓ | × | × |
| 2023 [19] | VCG auction theory | Full | × | ✓ | × | × |
| **Our proposal** | **Game theory, RL, state sharding** | Partial | ✓ | ✓ | ✓ | ✓ |

can lead to inefficiencies due to offloading costs, BC gas fees, and redundancy. To make BC sustainable for offloading, we need methods to balance memory, cost, and trust, particularly in the metaverse, where subtasks are prominent.

In the context of computational offloading, our focus is on state sharding, specifically investigating the optimal Blockchain Redundancy Factor (BRF) for a defined user group engaged in partial offloading. Notably, our approach diverges from others [27, 28] as we do not aim to alter the consensus algorithm. Instead, like [33-37] schemes, our emphasis lies in optimizing data replication, considering dynamic factors such as cross-shared communication overhead, shard allocation, BC pricing, and the overall cost of offloading to assist users in making an optimal offloading decision.

Motivated by these considerations, we consider the dynamic redundancy-aware BC-based partial computation offloading (PCO) for the metaverse in COIN framework. Specifically, the BRF (data replica) is updated periodically based on the user computation demand, cost, and price constraints to assist users in PCO. With the assistance of an optimal BRF, users can make an informed decision to execute their tasks locally or remotely. The main contributions of the paper are summarised as follows:

- We formulate a joint BRF and PCO problem in a multichannel wireless environment to minimize the offloading cost of user equipment over each time slot. It is intractable to determine its optimal solution due to the dynamic BC pricing, computation overhead, and complex wireless access in multiuser PCO scenarios. We theoretically prove that the problem is NP-hard using the maximum cardinality bin packing problem.

- To solve this problem, we decomposed the problem into two subproblems: the PCO problem from the user perspective and the BRF problem from the BC or network side. The PCO problem is reformulated into a multiuser PCO game and is proven to have Nash equilibrium (NE) with a convergence guarantee.

- For the second subproblem, the BRF, a double deep Q-network (DDQN), is proposed to learn and predict the optimal redundancy factor over time to assist players in maximizing their utility under unknown user task request information.

- We conducted a numerical simulation to evaluate the BRF-assisted PCO approach. The investigation results suggest that the proposed approach significantly reduces the computation cost and storage pressure on nodes while maintaining an optimal tradeoff between BC decentralization, communication or computation overhead, transaction prices, and mining incentives.

### C. Organization

The paper is organized as follows. Section II presents the system model, including the various system models and the joint PCO and BRF problem formulation. Next, Section III proposes an efficient scheme to solve the original problem using game theory and the DDQN. Section IV presents the numerical investigations to verify the efficacy of the proposed scheme, and Section V concludes the paper.

## II. SYSTEM MODEL

In this study, we considered XR applications to be a constituent of the metaverse, with growing interest across a spectrum of users. As highlighted in Fig.1, the XR processing entails eight functionalities, grouped into four components: object tracking and detection, simultaneous localization and mapping and map optimization with a point cloud dataset, hand gesture and pose estimation, and multimedia processing and transport (e.g., rendering, encoding, etc.) [12]. Fig. 2(a) illustrates metaverse task offloading in network computing, where users generate tasks divided into subtasks for local execution or offloading to edge COIN nodes (EINs) or fog COIN nodes (FINs). In contrast to previous studies[2, 38-43] addressing atomic tasks, our focus is on divisible tasks in the metaverse. The critical notation used in this article is summarised in Table II.

Let $\mathcal{K} = \{1, 2, \ldots, K\}$ represent user index set. We assumed each metaverse task $f \in \mathcal{F}$ is denoted by the tuple parameters: $\langle I_f, V_f, P_f \rangle$ where $I_f, V_f,$ and $P_f$ denote task $f$ input size; software volume, and computation load, respectively. Users, denoted by $k$, are assigned tasks consisting of subtasks task $f = \{f_{k,0}, f_{k,1}, f_{k,2} \ldots f_{k,j}\}$, where $j$ is the number of subtasks. Fig. 2(b) depicts the system operational mechanism. Users generate these tasks and request their local or remote execution (either FIN or EIN) at the beginning of each time slot $t$. Similar to [2, 44], we assumed that each subtask must be accomplished within

TABLE II

## NOTATION SUMMARY

| Notation | Definition |
|---|---|
| $K, F, M$ | Number of users, number of tasks, number of subchannels |
| $\mathcal{K}, \mathcal{F}, \mathcal{M}, \mathcal{A}$ | User set, task set, subchannel set, access point set |
| $L, FIN, EIN$ | Local computing, fog COIN, edge COIN |
| $I_f, V_f, P_f$ | Task $f$ input size, task $f$ software volume, task $f$ computation load |
| $\omega_{k,t}$ | Data transmission rate of user $k$ in slot $t$ |
| $\rho_k, F_k^L, F_k^{FIN}, F_k^{EIN}$ | User $k$ transmission power, user $k$ CPU, FIN computing capacity, EIN computing capacity |
| $\delta_{FIN}, \delta_{EIN}$ | FIN cache size, EIN cache size |
| $\mathcal{S}_{k,t}$ | User $k$ TS decision in slot $t$ |
| $b_f^{(t)}$ | Offloading destination for task $f$ in slot $t$ |
| $\mathcal{R}_{k,t}$ | User $k$ received interference in slot $t$ |

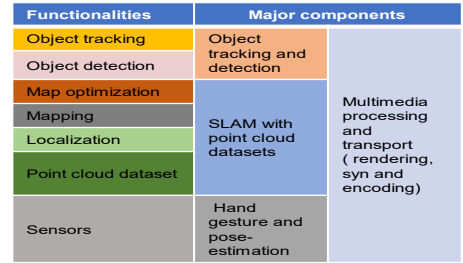

**Fig. 1.** Typical extended reality (XR) task processing on a device.

the time slot $t$. The COIN controller updates redundancy factors and strategies at the end of each time slot to assist users in PCO during the next time slot.

Let $\mu_t = \left\{\mu_1^{(t)}, \mu_2^{(t)}, \ldots, \mu_k^{(t)}\right\}|$ represent the user requests at time slot $t$. For each user $k$, the task request state is given by $\mu_k^{(t)} \in \bar{\mathcal{F}} = (\bar{\mathcal{F}}\{0\} \cup \mathcal{F})$, where $\mu_k^{(t)}$=0 indicates no request, and $\mu_k^{(t)} = f$ ($f \in \mathcal{F}$) indicates a request for task $f$. Relying on the Markov chain with an unspecified transition probability, $\mu_k^{(t)}(\forall k \in K)$ in time slot $(t + 1)$ only relies on the requests in slot $t$ with the number of options $(F + 1)$.

### A. Data Redundancy

Blockchain scalability involves maintaining essential functions as BC data volume grows without compromising BC characteristics. Inspired by previous work [21], which advocates for minimal redundancy, we explored the optimal redundancy level to maintain functional equivalence among nodes and preserve BC decentralization. We introduced the Blockchain Redundancy Factor (BRF), allowing users (or the system) to choose their desired redundancy level via a variable called the redundancy factor, $r$. Thus, we defined the BRF for computational offloading as follows.

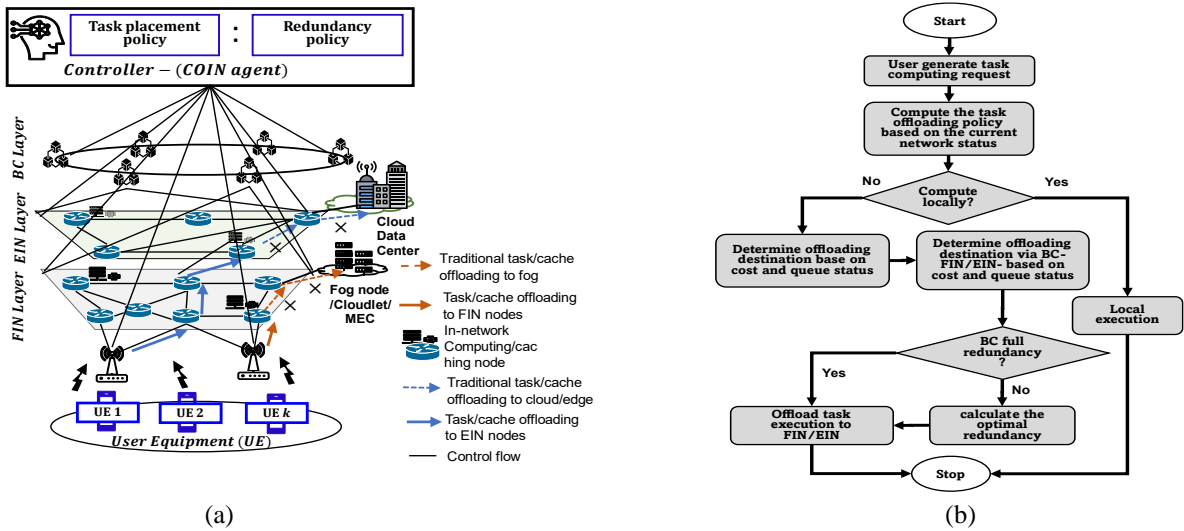

(a)

(b)

**Fig. 2.** Metaverse task offloading in computing in the network (COIN): (a) Network structure where COIN agents proactively select the blockchain redundancy factor (BRF) and user equipment can execute the task in three modes; (b) Operation mechanism flowchart in a time slot.

**Definition 1 (BRF):** *The BRF of offloading a given computation subtask $j$ via transaction as the $i$th block is the number of nodes that store data, denoted as the redundancy factor $r_f^{(t)}$. The average redundancy $r_{av}$ is the mean of redundancy factor across all blocks for the subtasks of use set $k$. The maximum redundancy factor $r_{max}$ and minimum redundancy factor $r_{min}$ refer to the maximum and minimum values among all redundancy blocks, respectively. Thus, the redundancy factor is formally defined as $r_f^{(t)} \in [r_{min}, r_{max}]$.*

Let $\beta_f^{(t)} \in \{0, 1\}$ denote the redundancy state of task $f$ in time slot $t$, where $\beta_f^{(t)} = 1$ for a redundancy-aware case ($1 \leq r < r_{max}$); otherwise, $\beta_f^{(t)} = 0$, indicating the nonredundancy-aware case ($r = r_{max}$). Thus, in given time slot $t$, the redundancy state is characterized by $\boldsymbol{\beta_t} = \left\{\beta_1^{(t)}, \beta_2^{(t)}, ..., \beta_F^{(t)}\right\}$. At any given time, $t$, the BRF decision profile is $\boldsymbol{r_t} = \left\{r_1^{(t)}, r_2^{(t)}, ..., r_F^{(t)}\right\}$. Due to constraints on the maximum redundancy level ($r_{max}$) and the total available BC offloading price ($P_{BC}$) for offloading via BC, the redundancy state must satisfy the following:

$$\sum_{f \in \mathcal{F}} \mathfrak{u} \cdot \left(\beta_f^{(t)} + r_f^{(t)}\right) \cdot V_f \leq P_{BC} \qquad (1)$$

$$\beta_f^{(t)} + r_f^{(t)} \leq r_{max} \qquad (2)$$

where $\mathfrak{u}$ is the price cost per MB offloaded, and 1 and 2 specify the maximum redundancy and user spending capacity contraints.

### B. Communication Model

In this scenario, we consider wireless uplink communication, where the access point $a \in \mathcal{A}$ is shared equally among the set of connected nodes $\mathcal{N}_a$. Thus, the uplink rate for remote

offloading for node $k$ in time $t$ is [2, 45]:

$$\omega_{k,t} = \frac{B}{M} log \left(1 + \frac{\rho_k \eta_k}{\sum_{n \in \mathcal{K} \setminus \{k\}, s_{n,t} = s_{k,t}} \rho_n \eta_n + \sigma^2}\right) \qquad (3)$$

where $\rho_k$ denotes the transmission power for user $k$, $\eta_k$ is the channel gain, $\sigma^2$ indicates the variance of complex white Gaussian channel noise, and $\sum_{n \in \mathcal{K} \setminus \{k\}, a_{n,t} = a_{k,t}} \rho_n \eta_n$ is the interference of user $k$ induced by other users. The values for $\rho_k$ and $\eta_k$ depends on the offloading destination (FIN or EIN). Users experience increased interference and reduced transmission rates when multiple users offload through the same channel, leading to higher energy consumption and offloading costs.

### C. Secured Offloading via the Blockchain

BC ensures data integrity and consistency through Byzantine fault tolerance-based consensus during secure sharing or offloading. During secure offloading via the BC, latency involves data transmission, computation, and transaction verification [16]. The latency depends on essential factors: the communication rate between BC nodes ($\omega_{k,t}$), the computing frequency of the primary BC node ($F_k$.), and the computing requirements for signature and message validation ($\mathcal{B}$ and $\delta$). The latency incurred when the tasks are remotely executed via the BC under a given BRF $r$ includes broadcast, pre-prepare, prepare, commit, and reply step latency:

1. Broadcast: When offloading to the FIN or EIN, tasks are sent to other BC nodes, with the primary node receiving and verifying each task as a transaction. The overall latency for this step includes both the transmission latency for offloading and the computing latency for transaction validation by the primary node. Hence, the latency is described as follows:

$$T_{k,f}^{bd} = \frac{I_f + V_f}{\omega_{k,t}} + \frac{rQ^b(\mathcal{B} + \delta)}{F_k(t)} \qquad (4)$$

where $Q^b$ denotes the number of transactions in a new block.

2. Pre-prepare: Upon transaction verification, the primary node generates a new signed block, which is transmitted to other nodes for consensus. The pre-prepare latency comprises transmission and computation, including block delivery and validation. The latency of this step is as follows:

$$T_{k,f}^{prep} = \frac{rQ^b(l_f + V_f)}{\omega_{k,t}} + \frac{\mathcal{B}(m-1)\delta}{F_k(t)} + \max_{Ej \in \mathcal{C}} \left\{ \frac{(\mathcal{B}+\delta)(Q^b+1)}{F_j(t)} \right\} \quad (5)$$

where $m$ represents the number of EINs.

3. Prepare: After verification, participating nodes add their signatures and verification codes to the results, broadcasting them to others. Each node cross-verifies the results with information from at least $2n$ sources, where $n$ denotes the number of abnormal participants. The latency for the signature and verification of this step is expressed as

$$T_{k,f}^{pre} = \frac{rQ^b(l_f + V_f)}{\omega_{k,t}} + \max_{Ej \in \mathcal{C}} \left\{ \frac{2n(\mathcal{B}+\delta)+\mathcal{B}+(m-1)\delta}{F_j(t)} \right\}. \quad (6)$$

4. Commit: In the commit process, each node generates a commit message, adds a signature, and broadcasts it to other nodes, cross-verifying with at least $2n$ sources, as in the pre-prepare step. If the $2n$ value is not achieved within a time threshold, the commit process is aborted, and the next block generation process begins. The commit delay is expressed as follows:

$$T_{k,f}^{com} = \frac{rQ^b(l_f + V_f)}{\omega_{k,t}} + \max_{Ej \in \mathcal{C}} \left\{ \frac{2n(\mathcal{B}+\delta)+\mathcal{B}+(m-1)\delta}{F_j(t)} \right\}. \quad (7)$$

5. Reply: After the commit process, each node adds its signature and code to the results, which are then sent to the primary node. The primary node validates these messages with at least $2n$ others before adding the new block to the chain. The expected delay in the reply phase is

$$T_{k,f}^{rpl} = \frac{rQ^b(l_f + V_f)}{\omega_{k,t}} + \max_{Ej \in \mathcal{C}} \left\{ \frac{Q^b(\mathcal{B}+\delta)}{F_j(t)} \right\} + \frac{2nQ^b(\mathcal{B}+\delta)}{F_k(t)}. \quad (8)$$

The overall latency for secure offloading according to the above analysis is given as follows:

$$T_{k,f}^{BC}(t) = T_{k,f}^{bd} + T_{k,f}^{prep} + T_{k,f}^{pre} + T_{k,f}^{com} + T_{k,f}^{rpl}. \quad (9)$$

### D. Computation Model

We considered local computation by the UE, FIN, and EIN. In the following, we elaborate on the three computation destinations:

#### 1) Local Computing (LIN)

The time taken to execute a given task $\langle l_f, V_f, P_f \rangle$ in LIN only includes the processing time on the local node at time $t$ and is defined as:

$$T_{k,f}^L = \frac{P_f}{F_k^L(t)} \quad (10)$$
$$\text{s.t } F_k^L(t) \leq F_{k,max}$$

where $F_i^L$ is the actual frequency at which a local node can execute the task. The corresponding energy consumption of executing locally at a frequency $F_i^L$ is linearly proportional to the square of the $F_k^L(t)$ and is given as

$$\mathcal{E}_t^L = \tau \left( F_k^L(t) \right)^2 P_f \quad (11)$$

where $\lambda \sim 10^{-11}$ is the energy coefficient of the user device determined by the chip architecture [41, 46]. Similar to [40, 42], the cost model for user $k$ is modeled in terms of the completion time and energy consumption as

$$\mathcal{C}_{k,f,t}^L = f(\delta_k^T T_k^L, \delta_k^E \mathcal{E}_k^L) = \delta_k^T T_k^L + \delta_k^E \mathcal{E}_k^L, \quad (12)$$

where $\delta_k^T$ and $\delta_k^E$ are the delay and energy preference (weight), respectively, and $0 \leq \delta_k^T, \delta_k^E \leq 1$, $\delta_k^T + \delta_k^E = 1$.

#### 2) Fog COIN (FIN)

When offloading a task of input $\langle l_f, V_f, P_f \rangle$ to remote computing nodes, such as the FIN, the total delay $T_{k,f,t}^{FIN}$ consists of three parts: data transmission time $T_{k,f}^t(t)$, task execution time $T_{k,F}^{exe}(t)$, and blockchain-related delay $T_k^{BC}$. The $T_{k,f,t}^{FIN}$ is expressed as follows:

$$T_{k,f,t}^{FIN} = T_{k,f}^t(t) + T_{k,F}^{exe}(t) + T_k^{BC}$$
$$= \frac{l_f + V_f}{\omega_{k,t}} + \frac{P_f}{F_k^{FIN}} + T_k^{BC} \quad (13)$$

where $F_k^{FIN}$ is the actual frequency at which the FIN can execute a task, and $\omega_{k,t}$ is given by (3). The results retrieval delay is considered negligible compared to the input data $l_f$ [47].

The PCO problem aims to minimize overall system cost. We employed the M/M/1 queuing model, which includes waiting time and utilization delay at the FIN. Waiting time is the duration subtasks spent in the queue before execution, and utilization delay accounts for the impact of the utilization factor on queue delay. The waiting time is the sum of remaining service times for queued subtasks and the current subtask's service time. A utilization factor of 1 indicates full capacity, resulting in the full remaining service time as queue delay; a factor below 1 implies lower capacity. The queue model is described as follows.

We let $q_{FIN}$ denote the FIN queue buffer. From service time $s_{FIN}(t) = \frac{P_f}{F_k^{FIN}} + \frac{l_f + V_f}{\omega_{k,t}}$, the arrival time is $a_{FIN}(t) = \frac{1}{s_{FIN}(t)}$. From the FIN rate $r_{FIN}(t) = \begin{cases} \frac{1}{\sum_j q_{FIN}(j) + s_{FIN}(t)}, & \text{if } q_{FIN} > 0 \\ \infty, & \text{otherwise} \end{cases}$, the utilization factor $u_{FIN} = \begin{cases} \frac{a_{FIN}(t)}{r_{FIN}(t)}, & \text{if } r_{FIN}(t) \neq \infty \\ 1, & \text{otherwise} \end{cases}$. The queue delay is calculated as

$$Q_{f,t}^{FIN} = \begin{cases} \sum_{j \subseteq f} q_{FIN}(j) + s_{FIN}(t), & \text{if } u_{FIN} = 1 \\ \frac{u_{FIN}^2}{1 - u_{FIN}} s_{FIN}(t), & \text{otherwise} \end{cases} \quad (14)$$

In this scenario, the cumulative latency cost associated with offloading to the FIN includes processing time, transmission time, and queuing delay:

$$T_{k,f,t}^{FIN} = \frac{P_f}{F_k^{FIN}} + \frac{l_f + V_f}{\omega_{k,t}} + T_k^{BC} + Q_{f,t}^{FIN} \quad (15)$$

where the first three terms on the right-hand side of (15) are computed by (13).

For offloading a given task to the FIN, the energy consumption is characterized by the energy used in offloading the task input size of a given task $I_f$, with negligible energy usage for connection scanning [40]. The energy consumption can be expressed as

$$\mathcal{E}_{k,f,t}^{FIN} = \rho_k \frac{I_f + V_f}{\omega_{k,t}} \quad (16)$$

where $\rho_k$ is the transmission power of the device, and $\omega_{k,t}$ is defined in (1). The corresponding total cost for the user $k$ considering the delay and energy, is

$$\mathcal{C}_{k,f,t}^{FIN} = f\left(\delta_k^T T_{k,f,t}^{FIN}, \delta_k^E \mathcal{E}_{k,f,t}^{FIN}\right) = \delta_k^T T_{k,f,t}^{FIN} + \delta_k^E \mathcal{E}_{k,f,t}^{FIN} \quad (17)$$

where $\delta_k^T$ and $\delta_k^E$ are the delay and energy weight, respectively, as expressed in (12).

### 3) Edge COIN (EIN)

For simplicity, we assume the EIN has sufficient resources to compute the offloaded task, $I_f$. Its total delay $T_{k,f,t}^{EIN}$ consists of transmission delay $T_{k,EIN}^t$, processing delay $T_{k,EIN}^{exe}$, and BC-related delay $T_k^{BC}(t)$. Similar to FIN, the return delay is negligible [47]. The $T_{k,f,t}^{EIN}$ is express as

$$\begin{aligned} T_{k,f,t}^{EIN} &= T_{k,EIN}^{exe} + T_{k,EIN}^t + T_k^{BC}(t) \\ &= \frac{I_f + V_f}{\omega_{k,t}} + \frac{P_f}{F_k^{EIN}} + T_k^{BC}(t) \end{aligned} \quad (18)$$

where $F_k^{EIN}$ is the actual frequency for the EIN and $\omega_{k,t}$ is given by (3).

Similar to the FIN, we model the additional delay at EIN for the task offloading problem using the M/M/1 queueing model, which considers waiting time and delay due to utilization. We let $q_{EIN}$ denote the FIN queue buffer. From service time $s_{EIN}(t) = \frac{P_f}{F_k^{EIN}} + \frac{I_f + V_f}{\omega_{k,t}}$, the arrival time $a_{EIN}(t) = \frac{1}{s_{EIN}(t)}$. From the EIN rate $r_{EIN}(t) = \begin{cases} \frac{1}{\sum_j q_{EIN}(j) + s_{EIN}(t)}, & if\ q_{EIN} > 0 \\ \infty, & otherwise \end{cases}$, the utilization factor $u_{EIN} = \begin{cases} \frac{a_{EIN}(t)}{r_{EIN}(t)}, & if\ r_{EIN}(t) \neq \infty. \\ 1, & otherwise \end{cases}$ The queue delay is calculated as

$$Q_{f,t}^{EIN} = \begin{cases} \sum_{j \subseteq f} q_{EIN}(j) + s_{EIN}(t), & if\ u_{EIN} = 1 \\ \frac{u_{EIN}^2}{1 - u_{EIN}} s_{EIN}(t), & otherwise. \end{cases} \quad (19)$$

For the case of the EIN, the total offloading latency is the summation of the processing, transmission time, and queuing delay:

$$T_{k,f,t}^{EIN} = \frac{P_f}{F_k^{EIN}} + \frac{I_f + V_f}{\omega_{k,t}} + T_i^{BC}(t) + Q_{f,t}^{EIN} \quad (20)$$

where the first three terms on the right-hand side of (20) are computed by (18). Likewise, the EIN must complete the task within the current time slot $t$, so the delay transmission is modeled as follows:

$$\mathcal{E}_{k,f,t}^{EIN} = \rho_k \frac{I_f + V_f}{\omega_{k,t}}. \quad (21)$$

The total cost of offloading a task $f$ to the EIN is

$$\mathcal{C}_{k,f,t}^{EIN} = f\left(\delta_k^T T_{k,f,t}^{EIN}, \delta_k^E \mathcal{E}_{k,f,t}^{EIN}\right) = \delta_k^T T_{k,f,t}^{EIN} + \delta_k^E \mathcal{E}_{k,f,t}^{EIN}. \quad (22)$$

### E. Problem Formulation

This work proposes a redundancy-aware mechanism that assists PCO. The COIN controller (can be a service) proactively determines redundancy factors and offers computing services in the next slot. Users decide on partial offloading policies at each time slot, executing tasks locally or offloading them to the FIN or EIN via the BC. Tasks offloaded via the BC can operate in nonredundancy-aware (full redundancy) or redundancy-aware (partial redundancy) modes. In the nonredundancy-aware mode, subtasks are replicated across all BC nodes, increasing offloading. Conversely, the redundancy-aware mode reduces data redundancy, storing data partially in some nodes to lower replicas and related costs. The costs of nonredundancy-aware ($\mathcal{C}_{k,f,t}^O$) and redundancy-aware offloading ($\mathcal{C}_{k,f,t}^R$) are as follows:

$$\mathcal{C}_{k,f,t}^O = \mathbb{1}\left(r_f^{(t)} = r_{max}\right)\left\{\left(1 - b_f^{(t)}\right)\mathcal{C}_{k,f,t}^{FIN} + b_f^{(t)}\mathcal{C}_{k,f,t}^{EIN}\right\} \quad (23)$$

$$\mathcal{C}_{k,f,t}^R = \mathbb{1}\left(r_f^{(t)} \in (r_{min}, r_{max})\right)\left\{\left(1 - b_f^{(t)}\right)\mathcal{C}_{k,f,t}^{FIN} + b_f^{(t)}\mathcal{C}_{k,f,t}^{EIN}\right\} \quad (24)$$

where $\mathbb{1}(.)$ is an indicator function, which is 1 when the parentheses are valid; otherwise, it is 0. For ease of representation, the corresponding delay and energy consumption of the nonredundancy-aware and redundancy-aware offloading modes are denoted as $T_{k,f,t}^O, \mathcal{E}_{k,f,t}^O$ and $T_{k,f,t}^R, \mathcal{E}_{k,f,t}^R$, respectively.

We aimed to minimize the average task execution cost of all users in an in-network over each time slot by jointly optimizing the PCO decisions and BRF policy. Thus, the cost of user $k$ at the time $t$ for task $i$ is formulated as

$$\mathcal{C}_{k,t} = \sum_f \mathbb{1}\left(\mathcal{S}_{k,t} = 0\right)\mathcal{C}_{k,f,t}^L + \mathbb{1}\left(\mathcal{S}_{k,t} \in \mathcal{M}\right) \left(\left(1 - \beta_f^{(t)}\right)\mathcal{C}_{k,f,t}^O + \beta_f^{(t)}\mathcal{C}_{k,f,t}^R\right) \quad (25)$$

where $\mathbb{1}(.)$ is an indicator function, which is 1 when the parentheses are valid and 0 otherwise. Equation (25) corresponds to three distinct cases. The first scenario involves the local execution (i.e., $\mathcal{S}_{k,t} = 0$) of task $f$. The total execution cost equals the local execution cost (i.e., $\mathcal{C}_{k,t} = \mathcal{C}_{k,f,t}^L$). In the second case, user $k$ chooses to offload the task to the FIN (i.e., $\mathcal{S}_{k,t} \in \mathcal{M}$ and $b_f^{(t)} = 0$), and the corresponding cost, $\mathcal{C}_{k,t} = \mathcal{C}_{k,f,t}^{FIN}$, comprises transmission and execution expenses. Likewise, in the third case, where the task is offloaded to the EIN ($\mathcal{S}_{k,t} \in \mathcal{M}$ and $b_f^{(t)} = 1$), the total cost, $\mathcal{C}_{k,t} = \mathcal{C}_{k,f,t}^{EIN}$, encompasses transmission and execution costs. Considering all these considerations, we can formulate the joint problem as follows:

$$\mathcal{J}_p : \min_{\mathcal{S}_t, r_t} \text{Lim}_{T \to \infty} \frac{1}{T}\sum_{t=1}^T \sum_{k \in \mathcal{K}} \mathcal{C}_{k,t} \quad (26)$$

$$s.t. \quad \sum_{v \in \mathcal{M} \cup \{0\}} \mathcal{S}_{k,v,t} \geq 1, \quad \forall k \in \mathcal{K}, v \in \mathcal{V}_k \quad (26a)$$

$$b_f^{(t)} \in \{0,1\} \quad (26b)$$

$$\sum_{f \in \mathcal{F}} \mathrm{u} \cdot \left(\beta_f^{(t)} + r_f^{(t)}\right) \cdot V_f \leq P_{BC}, \forall t \in \mathcal{T}, \quad (26c)$$

$$\beta_f^{(t)} + r_f^{(t)} \leq r_{max}, \forall f \in \mathcal{F}, \quad \forall t \in \mathcal{T}, \quad (26d)$$

$$\beta_f^{(t+1)} = \beta_f^{(t)} + r_f^{(t)}, \quad \forall t \in \mathcal{T}, \quad (26e)$$

$$(1 - b_f^{(t)}) T_{k,v,t}^O + b_f^{(t)} T_{k,v,t}^R = \mu \quad (26f)$$

$$\sum_{f \in} b_f^{(t)} V_f \le \delta_{EIN},$$

$$\sum_f (1 - b_f^{(t)}) V_f \le \delta_{FIN}, \forall t \in \mathcal{T}, \quad (26g)$$

$$\mathcal{S}_{k,t} \in \{0, 1I, M\} \quad \forall k \in \mathcal{K}, \forall t \in \mathcal{T}, v \in \mathcal{V}_k. \quad (26h)$$

Constraint (26a) indicates at least one subtask that is partially offloaded. Constraint (26b) signifies that the subtask can be executed remotely at either the FIN or EIN. Constraint (26c) implies the price cost constrain of offloading via the BC. Constraint (26d) enforces the maximum redundancy constraint, and (26e) reveals the BRF regulations. Constraint (26f) corresponds to the users' task execution delay constraint, and (26g) indicates the EIN and FIN cache size restriction, respectively. Constraint (26h) represents the task computing modes, where $\mathcal{S}_{k,t} = 0$ indicates the task is executed locally and $\mathcal{S}_{k,t} = m \ (m \in \mathcal{M})$ users offload the task to the FIN or EIN (redundancy-aware based offloading if $\beta_f^{(t)} = 1$ and nonredundancy-aware based if $\beta_f^{(t)} = 0$ through channel $m$). In addition, $\mu$ is the subtask dateline, and $\delta_F$ and $\delta_E$ are the cache memory of the FIN and EIN, respectively.

**Lemma 1.** *The problem $\mathcal{J}_p$ is NP-hard due to the interaction between PCO and BRF over different time slots.*

**Proof.** See Appendix A

## III. PARTIAL COMPUTATION OFFLOADING AND BLOCKCHAIN REDUNDANCY-AWARE UPDATE ALGORITHM

It is impossible to find an effective algorithm that can achieve an optimal solution for the problem $\mathcal{J}_p$ due to the interaction of the PCO and BRF over different time slots and the lack of user request transition probabilities. To address these problems, we split the problem into two subproblems: the PCO and BRF problems. First, the PCO problem is modeled as a multiuser ordinal potential game (OPG) from the user side, and a decentralized algorithm is proposed to find its Nash equilibrium (NE). Then, the BRF problem is modeled using a Markov decision process, and a DDQN is used to learn the optimal redundancy level policy.

### A. Multiuser PCO Algorithm

For each user, the deciding factor at any given time $t$ for the PCO decision $\mathcal{S}_{k,t}$ relies on the cost of computing $\mathcal{C}_{k,t}$. Additionally, $\mathcal{S}_{k,t}$ does not affect the redundancy factor decision in any time slot. Based on this, we focus on the PCO problem at a specific given time $t$ under any given redundancy state $\boldsymbol{\beta}_t$, and proposed a decentralized algorithm to attain the optimal PCO policy. Thus, we model the problem as a multiuser OPG as follows. We let $k$ PCO strategies be $\mathcal{S}_{k,t} = \{\mathcal{S}_{k,0,t}, \mathcal{S}_{k,1,t}, \dots, \mathcal{S}_{k,v,t} | \mathcal{S}_{k,v,t} = m \ (m \in M)\}$, where $v$ is the subtask number. All user strategies are defined as $\mathcal{S}_t = \{\mathcal{S}_{k,t} | \mathcal{S}_{k,t} \in \mathcal{S}_{k,t}, k \in \mathcal{K}\}$, where $\mathcal{S}_{k,t} = \mathcal{S}_{k,0,t} = 0$ denotes local subtask execution, and $\mathcal{S}_{k,t} = \mathcal{S}_{k,v,t} = m \ (m \in M)$ indicates the remote execution at the FIN ($b_f^{(t)} = 0$) or EIN ($b_f^{(t)} = 1$) node through channel $m$. The decomposed problem, the PCO problem, from problem $\mathcal{J}_p$ in slot $t$ is given as

$$\mathcal{J}_{p1} : \min_{\mathcal{S}_t} f_t(\mathcal{S}_t) = \sum_{k \in \mathcal{K}} \mathcal{C}_{k,t} \quad (27)$$

s.t $26a, 26b, 26f$, and $26h$ hold.

Similar to the work in [2], $\mathcal{S}_t$ of the PCO is a combinatorial problem with $(M + 1)$ value selections, which is challenging to solve over a multidimensional discrete space $\{0, 1, 2, \dots M\}^{K \times v}$. Thus, problem $\mathcal{J}_{p1}$ is transformed into a multiuser cooperative strategic game $\mathscr{g} = \langle \mathcal{K}, \{\mathcal{S}_{k,t}\}_{k \in \mathcal{K}}, f_t(\mathcal{S}_t) \rangle$, where $\mathcal{K}$, $\mathcal{S}_{k,t}$, and $f_t(\mathcal{S}_t)$ are the game player set, user $k$ strategy, and computing cost, respectively. The game's goal is to attain an NE solution $\mathcal{S}_t^* = \{\mathcal{S}_{1,t}^*, \mathcal{S}_{2,t}^*, \dots \mathcal{S}_{K,t}^*\}$, where no user can decrease its cost by changing its decision.

In this game $\mathscr{g}$, a given user remotely executes its subtasks if the local execution cost is larger than the subtask offloading cost (i.e., $\mathcal{C}_{k,t}^L \ge (1 - b_f^{(t)}) \mathcal{C}_{k,t}^F + b_f^{(t)} \mathcal{C}_{k,t}^E$). By substituting (12), (17), and (22) into the following inequality:

$$\delta_k^T T_k^L + \delta_k^E \mathcal{E}_k^L \ge \frac{(I_f + V_f)(\delta_k^T + \delta_k^E \rho_k)}{\omega_{k,t}} + \delta_k^T T_{k,FIN}^{exe}$$
$$- b_f^{(t)} (\delta_k^T T_{k,FIN}^{exe} - \delta_k^T T_{k,EIN}^{exe})$$
$$\omega_{k,t} \le \frac{(I_f + V_f)(\delta_k^T + \delta_k^E \rho_k)}{\delta_k^T (T_k^L - T_{k,FIN}^{exe}) + \delta_k^E \mathcal{E}_k^L + b_f^{(t)}(\delta_k^T T_{k,FIN}^{exe} - \delta_k^T T_{k,EIN}^{exe})},$$

user $k$'s interference $(\mathcal{R}_{k,t})$ is derived as

$$\mathcal{R}_{k,t} = \sum_{n \in \mathcal{K} \setminus \{k\}, a_{n,t} = a_{k,t}} \rho_n \eta_n \le$$
$$\frac{\rho_k \eta_k}{2^{\frac{M(I_f + V_f)(\delta_k^T + \delta_k^E \rho_k)}{B\left(\delta_k^T(T_k^L - T_{k,FIN}^{exe}) + \delta_k^E \mathcal{E}_k^L + b_f^{(t)}(\delta_k^T T_{k,FIN}^{exe} - \delta_k^T T_{k,EIN}^{exe})\right)}} - 1} - \sigma^2.$$
$$(28)$$

The inference threshold $(\lambda_k)$ for user $k$ is given as

$$\lambda_k = \frac{\rho_k \eta_k}{2^{\frac{M(I_f + V_f)(\delta_k^T + \delta_k^E \rho_k)}{B\left(\delta_k^T(T_k^L - T_{k,FIN}^{exe}) + \delta_k^E \mathcal{E}_k^L + b_f^{(t)}(\delta_k^T T_{k,FIN}^{exe} - \delta_k^T T_{k,EIN}^{exe})\right)}} - 1} - \sigma^2.$$
$$(29)$$

The user can reduce its computational cost through offloading; otherwise, it is accomplished through local computing using (29) because low interference suggests reduced cost. Similar to [45], the game $\mathscr{g}$ is an ordinal potential game with the potential function:

$$\phi(\mathcal{S}_t) = \frac{1}{2} \sum_k^K \sum_{n \ne k} \rho_k \eta_k \rho_n \eta_n \mathbb{1}(\mathcal{S}_{n,t} = \mathcal{S}_{k,t})$$
$$(\mathcal{S}_{k,t} > 0) + \sum_{k=1}^K \rho_k \eta_k \lambda_k \mathbb{1}(\mathcal{S}_{k,t} = 0). \quad (30)$$

Therefore, a user $k$ offloads a subtask when $\mathcal{R}_{k,t} \le \lambda_k$, otherwise, it is executed locally.

**Remark 1**. *The PCO game $\mathscr{g}$ with the potential function $\phi(\mathcal{S}_t)$ is an ordinal potential game that achieves NE in finite iteration numbers.*

**Proof.** See Appendix B.

*Algorithm 1* solves the multiuser PCO problem, ensuring mutual user satisfaction. The algorithm initializes offloading decisions based on computational efficiency. It computes local computing efficiency ($\mathcal{L}_L$) using the formula: $P_f \cdot F_k^L / \rho_{k,} \cdot V_f$. For remote offloading, it computes FIN efficiency ($\mathcal{L}_{FIN}$) as $P_f \cdot F_k^L / F_k^{FIN} \cdot \delta_{FIN}$ and EIN efficiency ($\mathcal{L}_{EIN}$) as $P_f \cdot F_k^L / F_k^{EIN} \cdot \delta_{EIN}$. This initialization results in faster convergence to optimal solutions. The algorithm computes each user's PCO by solving constraints 26a, 26b, 26f, and 26h in a repeat-until loop until the end message is received. Users calculate inference and transmission rates for each subtask, defining their strategy space. Constraints 26a, 26f, and 26h guide optimal PCO decisions, minimizing objective function $f_t(\mathcal{S}_{k,t}, \mathcal{S}_{-k,t})$ based on latency and energy consumption. Users can request updates to the PCO policy, maintaining decisions if no updates are received from the COIN controller. Upon receiving the END message, the users offload their tasks. We analyze the convergence behavior of Algorithm 1 in Lemma 2.

**Lemma 2.** The multiuser PCO game $\mathcal{g}$ can attain an NE solution within a finite iteration:
$$\frac{\frac{1}{2}\mathcal{K}^2\Omega_{max}^2 + K(\Omega_{max}\lambda_{max} - \Omega_{min}\lambda_{min})}{\pi\Omega_{min}} | \pi \in \mathbb{R}^+.$$

**Proof.** See Appendix C.

### B. Deep Reinforcement Learning-based Redundancy Factor Update Algorithm

Under any given redundancy, the state $\boldsymbol{\beta_t}$ and user demand at any given time slot, a mutually satisfactory PCO decision $\mathcal{S}_t^*$ can be obtained for all users through Algorithm 1. The problem $\mathcal{J}_p$ can be transformed into the BRF problem $\mathcal{J}_{p2}$ by substitute $\mathcal{S}_t^*$ in the problem as

$$\mathcal{J}_{p2} : \min_{\boldsymbol{r_t}} \operatorname*{Lim}_{T\to\infty} \frac{1}{T} \sum_{t=1}^{T} \sum_{k \in \mathcal{K}} \hat{\mathcal{C}}_{k,t} \quad (31)$$
$$\text{s.t. } 26c, 26d, 26e, \text{ and } 26g$$

where

$$\hat{\mathcal{C}}_{k,t} = \sum_f \mathbb{1}\big(\mathcal{S}_{k,t} = 0\big) \mathcal{C}_{k,f,t}^L + \mathbb{1}\big(\mathcal{S}_{k,t} \in \mathcal{M}\big)\Big(\big(1 - \beta_f^{(t)}\big)\mathcal{C}_{k,f,t}^O + \beta_f^{(t)}\mathcal{C}_{k,f,t}^R\Big). \quad (32)$$

With the redundancy factor decision $r_f^{(t)}$ depending on the BRF decision state $\boldsymbol{r_t}$ and the price of offloading via the BC $P_{BC}$ and incentive for mining $\mathcal{J}_{BC}$, it is challenging to obtain directly $\boldsymbol{r_t}$. Therefore, we first solve for the optimal redundancy state $\boldsymbol{\beta_{t+1}}$ in time slot $t + 1$, then solve for $\boldsymbol{r_t}$ based on $\beta_f^{(t)} + r_f^{(t)} = \beta_f^{(t+1)}$. The optimal redundancy state problem is formulated as follows:

$$\mathcal{J}_{p2} : \min_{\boldsymbol{\beta_{t+1}}} \operatorname*{Lim}_{T\to\infty} \frac{1}{T} \sum_{t=1}^{T} \sum_{k \in \mathcal{K}} \hat{\mathcal{C}}_{k,t} \quad (33)$$
$$\text{s.t } \sum_{f \in \mathcal{F}} \mathfrak{u} \cdot \big(\beta_f^{(t+1)} + r_f^{(t+1)}\big) \cdot V_f \leq P_{BC} \quad (33a)$$
$$\beta_f^{(t+1)} + r_f^{(t+1)} \leq r_{max}, \qquad r_{max} \in \mathbb{Z}^+ \quad (33b)$$
$$\beta_f^{(t+1)} \in \{0, 1\}, \quad (33c)$$
$$r_f^{(t+1)} \in [r_{min}, r_{max}],$$
$$1 \geq r_{min} < r_{max} | r_{min}, r_{max} \in \mathbb{Z}^+. \quad (33d)$$

1  **Initialization:** Each UE $k \in \mathcal{K}$ initializes its PCO decision by analyzing efficiency:
$$\mathcal{S}_{k,t} = \begin{cases} 0, & \text{if } \mathcal{L}_{FIN} < \mathcal{L}_L \\ 1, & \text{if } \mathcal{L}_{FIN} > \mathcal{L}_{EIN} \\ 2, & \text{otherwise} \end{cases}$$
2  **Repeat**
3      **For:** each UE $k \in \mathcal{K}$: **do**
4          **For:** each task subtask, $\upsilon \in \mathcal{V}_k$: **do**
5              Compute interference $\mathcal{R}_{k,t}$ and the transmission rate $\omega_{k,t}$
6              Compute the strategy $\mathcal{S}_{k,t}$ by solving constraint 26a, 26b, 26f, and 26h
7              Obtain the best optimal decision $\mathcal{S}_{k,t}^* = \arg\min_{\mathcal{S}_{k,t}} f_t(\mathcal{S}_{k,t}, \mathcal{S}_{-k,t})$
8              **if** $\mathcal{S}_{k,t}^* \neq \mathcal{S}_{k,t}$ **then**
9                  **Request** the controller to update the PCO policy
10                 **If** an optimal update is received,
11                     **then** update the PCO decision, $\mathcal{S}_{k,t} = \mathcal{S}_{k,t}^*$
12                 **end if**
13             **end if**
14         **end for**
15     **end for**
16 **Until** an END message is received
17 **Return** $\mathcal{S}_{k,t}$

Optimizing the redundancy factor $\boldsymbol{\beta_{t+1}}$ for user demand in time slot $t + 1$ involves selecting the most cost-effective decision among various redundancy factors $r_f^{(t)}$ while adhering to the BC price constraint. However, the BRF decision state $\boldsymbol{r_t}$ is determined at the time slot $t$, and the user demand and resource status are unknown due to unknown user request and resource status. The DDQN is employed to address this challenge by capturing the request model and predicting the subtask BRF decision state at the time slot $(t + 1)$ based on the system state of slot $t$. Thus, the problem is formulated as a Markov decision process, and the state, action, and reward are elaborated as follows:

**1) State:** In a multiuser scenario with diverse subtasks, the state of each subtask can vary due to factors like input size, computation requirements, software needs, and resource availability. Decisions made by other users also contribute to this complexity, resulting in an exponential increase of state space. To address this diversity and the state exploration challenge, we propose the following method for computing the system state for the users and their subtasks, ensuring effective representation of the environment: $S_t = S_{t,K} \in (F + 1)^K$, where $S_{t,k\in K} = \sum_{f=0}^{F} S_f$, $S_f = S_{\upsilon \subseteq f} = \chi_0 S_0 + \chi_1 S_1 + \chi_2 S_2 + \chi_3 S_3 | \chi_{f \in \{0,1,2,3\}} \in \{0,1\}$, and $S_{\upsilon \subseteq f}$ is the weighted sum of the user CPU and cache, subtask input parameters $\langle I_f, V_f, P_f \rangle$, FIN/EIN CPU, and cache size. We can derive a value that characterizes the state of each subtask for an individual user, and when aggregated, these individual states constitute the overall system state.

**2) Action:** The action in time slot $t$ is the redundancy state in $(t + 1)$ (i.e., $A_t = \boldsymbol{\beta_{t+1}} \in \{r_f, r_{max}\}^{F \times \upsilon}$).

**3) Reward**: The agent's goal is to minimize the cost while predicting the optimal BRF and maximizing the overall incentive $\mathcal{I}_{BC}$, considering the BC offloading price constraint $P_{BC}$. This creates a tradeoff between cost reduction and incentive maximization. Due to this tradeoff, we formulated the reward function to include the weighted sum of the cost difference (saving value) and incentive. The reward combines cost savings and incentives, with cost savings as $\mathcal{C}_{k,f,t}^{O} - \mathcal{C}_{k,f,t}^{R}$ and incentive as $\mathfrak{d}\,(I_f + V_f)$ where $\mathfrak{d}$ incentive per MB. The reward function is defined as follows:

$R_{t+1} = \mathfrak{w}_1(\sum_{k=1}^{K}\mathcal{C}_{t+1}^{O} - \sum_{k=1}^{K}\mathcal{C}_{t+1}^{R}) + \mathfrak{w}_2 \cdot \mathcal{I}_{BC}$

$R_{t+1} = \mathfrak{w}_1(\sum_{k=1}^{K}\mathcal{C}_{t+1}^{O} - \sum_{k=1}^{K}\mathcal{C}_{t+1}^{R}) + \mathfrak{w}_2\big(\mathfrak{d}\,(I_f + V_f)\,\big)$ (34)

where $\mathfrak{w}_1$ and $\mathfrak{w}_2$ are the weight of the cost difference and the incentive, respectively. For ease of training, we recommend standardizing the reward.

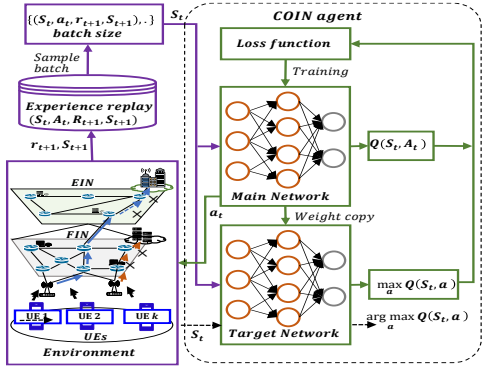

**Fig. 3.** Task offloading optimization using the double deep Q-network (DDQN).

The DDQN architecture employed in this study is presented in The DDQN architecture in this study, as shown in Fig. 3, utilizes the State Coding and Action Aggregate (SCAA) deep neural network (DNN) model [2]. The SCAA-DNN model is employed to handle the high-dimensional space associated with the complex action space of the problem. It consists of a primary network for learning Q-values and a target network for stability. The SCAA-DNN includes a dropout mechanism in the input layer and employs a two-layer architecture in the output layer to aggregate redundancy levels dynamically. This design reduces the impact of input order. Instead of $S_t$ for input, it uses $X_t = \{\mathbb{1}(S_{k,t} \in f)\,k \in K\}$.

Traditionally, the output state neurons of the DNN correspond to all possible actions for the subtasks with an output state action, $Q(S_t, A_t)$ However, it is impractical to iterate the network for each subtask individually due to multiple subtasks in the metaverse and the necessity to determine BRF actions for all of them within a single time slot. To this end, the SCAA-DNN consists of a neuron on the output layer $\Theta = (\Theta_1, \Theta_2 \ldots \Theta_v)$, representing the state-action value for all $v$ subtasks. The last layer has no activation unit; its output results from summing all input variables. The weight connection between the first and last layers is denoted as $w_L = (w_{1L}, w_{vL}, \ldots w_{VL})$. The state-action value for a specific action $A_t = b_{t+1}$ connected to $w_L$ is represented by $w_{fL} = b_v^{(t+1)}, \forall v \subseteq f \in \mathcal{F}$. Consequently, the DNN outputs the predicted state-action value $Q(S_t, A_t) = \sum_{v \subseteq f \in \mathcal{F}} b_v^{(t+1)}\Theta$.

The DDQN is trained using the ε-greedy policy at the end of time slot $t$; the COIN agent either decides offloading strategies based on the Q-values or randomly with a probability of ε. At the start of slot $(t + 1)$, users generate task-computing requests $\boldsymbol{S}_{t+1}$ and use *Algorithm 1* with the redundancy state $\boldsymbol{\beta}_{t+1}$ and system state to determine optimal PCO decisions $\boldsymbol{S}_{t+1}^{*}$. These decisions lead to offloading costs (i.e., $\mathcal{C}_{t+1}^{R}$). The redundancy state is temporarily set to empty (i.e., $\boldsymbol{\beta}_{t+1} = 0$), and the energy consumption is recorded as $\mathcal{C}_{t+1}^{O}$ to calculate the reward for the redundancy action $A_t = \boldsymbol{\beta}_{t+1}$. Experience memory (EM) stores the user request state $S_t$, action $A_t$, reward $R_{t+1}$, and subsequent time slot state $S_{t+1}$ from time slot $t$. Subsequently, DDQN selects a batch of data$(S_t, A_t, R_{t+1}, S_{t+1})$ from the EM for training purposes.

In the training phase, $A_t$ for $w_L$ is determined as $w_{fL} = \beta_v^{(t+1)}, \forall v \subseteq f \in \mathcal{F}$, with $X_t = \{\mathbb{1}(v_j^k \in f) : k \in K, j \in v\}$ as the input for processing. The prediction and approximation of the state-value action value are as follows:

$$Q(S_t, A_t) = R_{t+1} + \gamma\,\underset{a}{max}\,Q(S_{t+1}, a) \qquad (35)$$

where $\gamma \in (0,1)$ represents the discount factor. The target DNN is used to infer the value of $\underset{a}{max}\,Q(S_{t+1}, a)$. The main net handles action selection during the agent's interaction with the environment. The loss is computed using the Huber function to ensure stable learning, and the DNN is trained using the backward algorithm. The training algorithm is outlined in *Algorithm 2*.

At the inference phase, we determine the optimal BRF decision state in the time slot $(t + 1)$ using the following optimal offloading statement:

$\bar{\mathcal{J}}_{p2} : \underset{\boldsymbol{\beta}_{t+1}}{max} \sum_{f \in \mathcal{F}} \mathfrak{u} \cdot \big(\beta_f^{(t+1)} + r_f^{(t+1)}\big) \cdot V_f \; \Theta_v \qquad (36)$

$\text{s.t} \sum_{f \in \mathcal{F}} \mathfrak{u} \cdot \big(\beta_f^{(t+1)} + r_f^{(t+1)}\big) \cdot V_f \; \leq P_{BC} \qquad (36a)$

$\beta_f^{(t+1)} + r_f^{(t+1)} \leq r_{max}\,,\; r_{max} \epsilon\,\mathbb{Z}^{+} \qquad (36b)$

$\beta_f^{(t+1)} \epsilon\,\{0,1\}, \qquad (36c)$

$r_f^{(t+1)} \in\,[r_{min}, r_{max}],$

$\quad 1 \geq r_{min} < r_{max}|r_{min}, r_{max}\epsilon\,\mathbb{Z}^{+}. \qquad (36d)$

The problem $\bar{\mathcal{J}}_{p2}$ is a typical knapsack problem. Unlike [2, 48], which focuses on the software cache and partial offloading problems, we derive the solution for the BRF decision profile $\boldsymbol{r}_{t+1}$ using a recursive algorithm. We let $\in$ be an $F \times P_{BC}$ matrix in which $\in (v, P_{bc})$ denotes the optimal approach for carrying out subtask $v \subseteq f \in \mathcal{F}$ under the BC price $P_{bc}$ constraint. The recursive function $\in (v, P_{bc})$ can be expressed as follows:

| **ALGORITHM 2**: *DDQN Training Algorithm* |
|---|
| 1 | **Initialize:** Replay memory $EM$ to capacity $M$, the weight copy frequency ɡ |
| 2 | **Initialize:** The main DNN with random weight $\theta$. Copy the weight $\theta$ to the target DNN |
| 3 | **For:** time slot $t = 1:T$ **do** |
| 4 | With probability $\varepsilon$ select a random redundancy state $A_t$; otherwise, select $A_t = \arg\max_a Q(S_t, a)$ as a partial offloading state in slot $t + 1$. |
| 5 | Compute the reward $R_{t+1}$ using the $A_t$ in time in slot $t + 1$. |
| 6 | Store transition $(S_t, A_t, R_{t+1}, S_{t+1})$ in the $EM$ |
| 7 | Sample random minibatch $(S_t, A_t, R_{t+1}, S_{t+1})$ from the $EM$ |
| 8 | Based on the SCAA-DNN model, assign values to weights, $w_L$ |
| 9 | Assign $b_{t+1}$ to the weight of the TLA in the main network and obtain $Q(S_t A_t)$ |
| 10 | Using the loss function, perform gradient with respect to the DNN parameter. |
| 11 | Update the target network every ɡ slots. |
| 12 | **End for** |

$$\in (v, P_{bc}) = \arg\min_{a \in [r_{min}, r_{max}]} \big(\in (v - 1, P_{bc} - ų. (\beta_f^{(t+1)} + r_f^{(t+1)})V_f) + ų. (\beta_f^{(t+1)} + r_f^{(t+1)})V_f \Theta_v \big). \quad (37)$$

*Algorithm 3* presents the optimal redundancy state whose time complexity is $O(2FP_{BC} + F)$.

After obtaining the optimal redundancy state $\boldsymbol{r}_{t+1}$ in time slot $t + 1$, the COIN controller or agent can determine the optimal BRF policy for time slot $t$, (i.e., $r_f^{(t)} = \beta_f^{(t+1)} - \beta_f^{(t)}$). Following this, the COIN agent updates its BRF policy (data replica) and assists the PCO in the time slot $(t + 1)$. The step-by-step process of the DDQN inference phase is summarized in *Algorithm 4* for clarity. Additionally, to facilitate comprehension, Fig. 4 illustrates the interconnections between all algorithms and the physical system model.

| **ALGORITHM 3**: *Solving for the optimal action* |
|---|
| | Input: $\Theta_t, \{V_v \subseteq f, f \in \mathcal{F}\}$, ų |
| | Output: The optimal redundancy state $\boldsymbol{\beta}_{t+1}$ |
| 1 | $\boldsymbol{\beta}_{t+1} = [0]_F, \in = [0]_{F \times P_{BC}}, \in_r = [0]_{F \times P_{BC}}; r_{t+1} = [r_{min}, r_{max}]$ |
| 2 | for each $v \in [1, F]$: do |
| 3 | if $v < F$ then |
| 4 | for each $P_{bc} \in [1, P_{BC}]$: do |
| 5 | if $v == 1$ then |
| 6 | $\in_r (v, P_{bc}) = \mathbb{1}(V_v < P_{bc})$ |
| 7 | $\in (v, P_{bc}) = \in_r (v, P_{bc})\Theta_v$ |
| 8 | else |
| 9 | $\in_r (v, P_{bc}) = \arg\min_{a \in [r_{min}, r_{max}]} (\in (v - 1, c - aV_v) + a\Theta_v)$ |
| 10 | $\in (v, P_{bc}) = \in_r (v, c)\Theta_v + \in (v - 1, c - \in_r (v, c)V_v)$ |
| 11 | end if |
| 12 | end for |
| 13 | else |
| 14 | $\in_r (F, P_{BC}) = \arg\min_{a \in [r_{min}, r_{max}]} (\in (F - 1, P_{BC} - aV_F) + a\Theta_F)$ |
| 15 | $\in (F, P_{BC}) = \in_r (F, P_{BC})\Theta_F + \in (F - 1, P_{BC} - \in_r (F, P_{BC})V_F)$ |
| 16 | end if |
| 17 | end for |
| 18 | $\boldsymbol{\beta}_{t+1}(F) = \in_r (F, P_{BC})$ |
| 19 | for each $v = F - 1: -1: 1$ do |
| 20 | $\boldsymbol{\beta}_{t+1}(v) = \in_r (v, P_{BC} - \sum_{v+1 \leq j \leq F} ų(\boldsymbol{\beta}_{t+1}(j) + r_{t+1}(j)) * V_j)$ |
| 21 | end for |
| 22 | return $\boldsymbol{\beta}_{t+1}$ |

| **ALGORITHM 4** *DDQN inference algorithm* |
|---|
| 1 | Initialize the weights and parameters of the SCAA model. |
| 2 | Input $X_t = \{\mathbb{1}(\mathcal{S}_{k,t} \in f) k \in K\}$ into the first layer of the model, then forward it through the model, and output $\Theta = (\Theta_1, \Theta_2, \dots, \Theta_v)$. |
| 3 | Solve the optimal redundancy state in the next time slot using Algorithm 3. |
| 4 | Calculate the optimal redundancy update policy based on $\beta_f^{(t)} = \beta_f^{(t+1)} - r_f^{(t)}, \forall f \in \mathcal{F}$ |

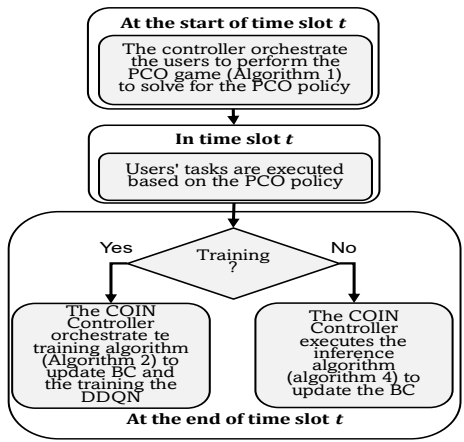

**Fig. 4.** Illustration of the interaction between the proposed algorithms and the system model.

## IV. NUMERICAL SIMULATIONS

This section presents the numerical simulation of the proposed system. We considered a scenario with $K$ users randomly distributed within $200 \times 200$ m cell region where the FIN and EIN are positioned at the center and edge of the cells, respectively (i.e., the FIN is closer to the user and equidistance to the EIN placed after the FIN). The energy coefficient is set as $5 \times 10^{-27}$ according to [49]. The channel gain is modeled as $\eta_k = \mathcal{P}_k(t)d_k^{-n}$ where $d_k$ is the distance between user $k$ and the remote COIN resource, $\mathcal{P}_k \sim \mathrm{Exp}(1)$ is exponentially distribution with a unit mean, representing the small-scale fading channel power gain from user $k$ to the COIN resource in slot $t$, and $n$ is the path loss factor [2]. For each task $f$ and user $k$, the input data size $I_f$ is uniformly and randomly selected from $[1, I_{jmax}]$ gigabytes. for subtasks $f_k = \{f_j : j \in J\}$. Similarly, software data size $V_f$ and CPU cycles for computing subtask $P_f$ are uniformly and randomly chosen from the intervals $[1, V_{jmax}]$ gigabytes and $[1, P_{jmax}]$ gigacycles, respectively. The system simulation was conducted using Python 3 on a Windows 10 core i5 system. Table III provides the simulation settings with default values unless otherwise specified.

To assess the effectiveness of our proposed model, we conducted evaluations against several common baselines, all of which are applicable to the proposed scenario involving task offloading through the BC under either full or partial redundancy. The baselines include OPG with full redundancy, OPG with a randomly assigned redundancy factor (partial redundancy), MEC under full redundancy, and random-based offloading:

- The OPG method considers a scenario where OPG offloading via the BC is based on full redundancy. The OPG has generally performed remarkably in solving task-offloading problems [50-53] and we included it as a benchmark to compare against our proposed redundancy-aware offloading approach.

TABLE III
SIMULATION SETTINGS

| Parameter | Value |
|---|---|
| User number: $K$ | 30 |
| Subtask number: $\mathcal{V}_k$ | 4 |
| Time slot: $T$ | 2000 |
| Wireless transmission bandwidth: $B$ | 50 MHz |
| Transmission power: $\rho_k^{EIN}/\rho_k^{FIN}$ | 0.5 W |
| Guassain noise variance: $\sigma^2$ | $2 \times 10^{-13}$ |
| CPU capability of user $k$: $F_k^L$ | 1 GHz |
| CPU capability of FIN: $F_k^{FIN}$ | 60 GHz |
| CPU capability of EIN: $F_k^{EIN}$ | 100 GHz |
| Cache size of FIN: $\delta_{FIN}$ | 3 GB |
| Cache size of EIN: $\delta_{FIN}$ | 5 GB |
| Number of channels: $M$ | 10 |
| DNN learning rate | 0.0008 |
| Experience replay memory size: $E$ | 10 000 |
| Batch size | 32 |
| Discount factor: $\gamma$ | 0.9 |
| $I_{imax}, V_{imax}, P_{max}$ | 5–10 |

- The OPG method assisted by a random redundancy factor (OPG-Rand) considers a scenario where OPG method offloading via the BC is based on a randomly assigned BRF unlike existing studies [50-53]. This baseline allows us to fairly evaluate the effectiveness of our proposed agent, which learns the optimal redundancy factor based on dynamic user demand and system status.

- In the MEC scenario, subtasks are computed locally or offloaded to MEC based on the caching memory available in MEC under the full redundancy scheme. The MEC method is considered to be equidistant with EIN to evaluate the benefit of COIN (reducing the edge's computing load) over MEC. This baseline enables us to observe the performance of the proposed redundancy-aware COIN-based approach against the state-of-the-art MEC baseline [15-20]. under full redundancy.

- Random computing is based on random partial offloading of subtasks to the FIN and EIN or locally computed on the device based on full redundancy. It serves as an intuitive baseline, as it is commonly used to evaluate state-of-the-art offloading methods via the BC.

To compare the proposed approach's performance fairly, we analyzed several aspects. Firstly, we assessed the average cost and reward across various training episodes in comparison to the baselines. Fig. 5(a) illustrates the average total system costs across different training episodes. The proposed model consistently had the lowest cost, with the random, MEC, OPG, and OPG-Rand models following closely. Notably, the proposed approach reduced costs by over 100% compared to the random approach. After around 1000 episodes, it maintained significantly lower costs. Additionally, Fig. 5(b) shows that the proposed method achieved a higher system reward, surpassing other methods by approximately 99%. We maintained 1000 epochs for subsequent investigation unless specified otherwise.

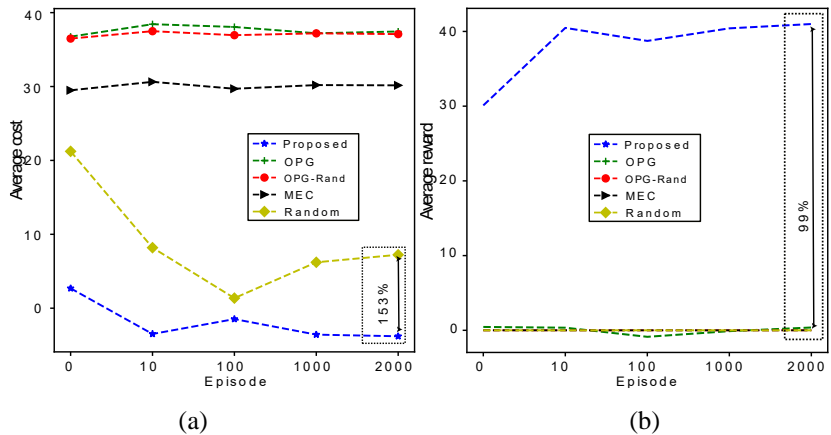

**Fig. 5.** Performance evaluation with respect to the iteration step under various episodes: (a) system cost and (b) system reward.

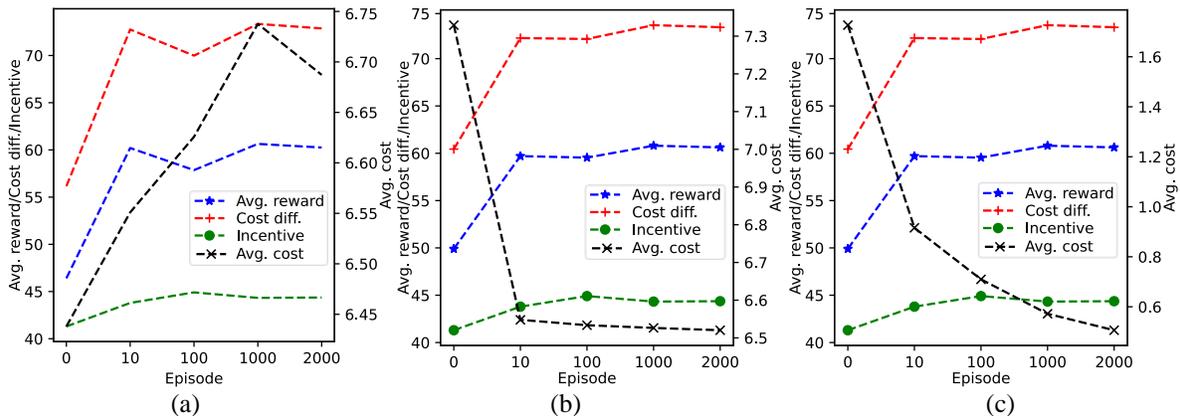

**Fig. 6.** Reward function evaluation against the average cost: (a) proposed agent, (b) OPG, and (c) OPG-Rand.

Furthermore, we analyzed the reward function in relation to average cost, aiming to minimize costs and maximize incentives. As explained in Section III-B, the reward comprises the weighted sum of the cost difference and incentive. The objective is to minimize system costs while maximizing overall incentives. The incentive is influenced by the redundancy factor, with partial redundancy potentially reducing competition and overall rewards. The agent's ability to learn optimal redundancy factor updates for cost reduction and increased rewards is evident in Fig. 6(a). Fig. 6(b) shows the OPG-based method, where the average cost continues to rise despite a widening cost difference. This occurs because OPG primarily focuses on cost reduction from the user's perspective with full redundancy, resulting in maximum incentives but higher cost overhead. As for the OPG-Rand approach in Fig. 6(c), the random assignment of the redundancy factor reduces the overall system cost effectively. This highlights the benefits of optimal redundancy factor allocation, particularly in reducing delays and overall system costs.

### B. Influence of Experimental Parameters on the System Model

#### 1) Influence of Computing Task Types

We evaluate the proposed system model's effectiveness by examining the impact of subtask types. We categorized computing subtasks as either data-intensive or compute-intensive, similar to [42]. Six computing tasks were considered, each comprising four subtasks (divisible tasks). The first three tasks (Tasks 1 to 3) were data-intensive, while the remaining three (Tasks 4 to 6) were compute-intensive. For data-intensive tasks, input size $I_f$ and software volume $V_f$ of the four subtasks are uniformly and randomly generated from {[10-20] MB, [0.5-2] GB}, respectively. The required CPU cycles $P_f$ for each data-intensive subtask were randomly chosen from {[1-4] gigacycles}. In the compute-intensive task type, $I_f$ and $V_f$ were generated uniformly and randomly from {[1-4] MB, [0.5-2] GB}, respectively, with corresponding $P_f$ randomly selected from {[5-20] gigacycles}.Some slight adjustments to the ranges of randomly generated values were made to accommodate variations in subtask sizes.

In terms of cost, our model outperformed all others, with reductions of 108% to 124% for data-intensive tasks (1–3) and 111% to 119% for compute-intensive tasks (4–6) compared to the second-best MEC model, as depicted in Fig. 7(a). Achieving very low latency is crucial in the metaverse to ensure the required quality of experience. Therefore, we assessed model performance based on latency. The proposed model achieved a latency reduction of 109% to 124% for data-intensive tasks and 111% to 119% for compute-intensive tasks compared to MEC (the second best), as illustrated in Fig. 7(b). In data-intensive tasks, latency primarily arises from communication transmission, while computational processes require additional latency. Irrespective of the task type, our COIN agent earned nearly 100% more average reward than the baselines, indicating its superior performance in cost reduction and reward maximization, as shown in Fig. 7(c)

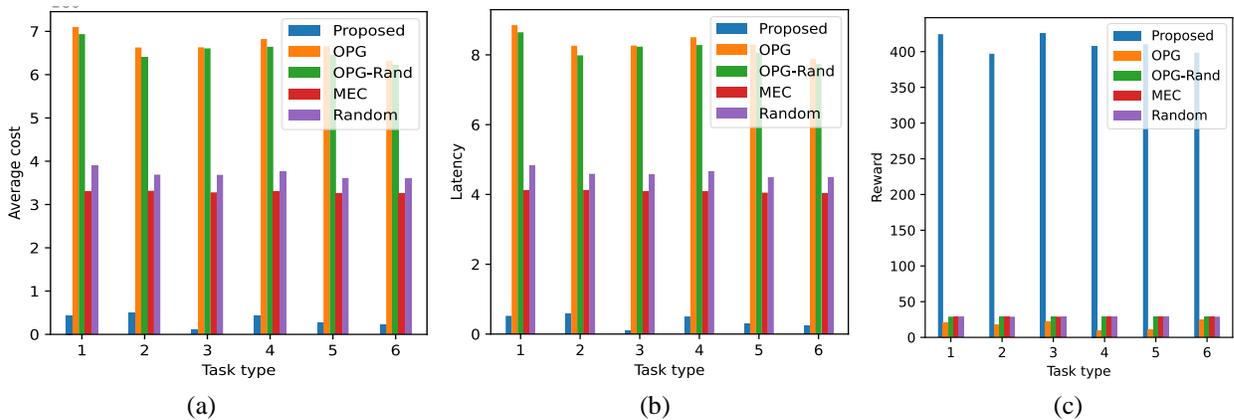

**Fig. 7.** Influence of the computing task types on the computation overhead: (a) average cost, (b) average latency, and (c) average reward.

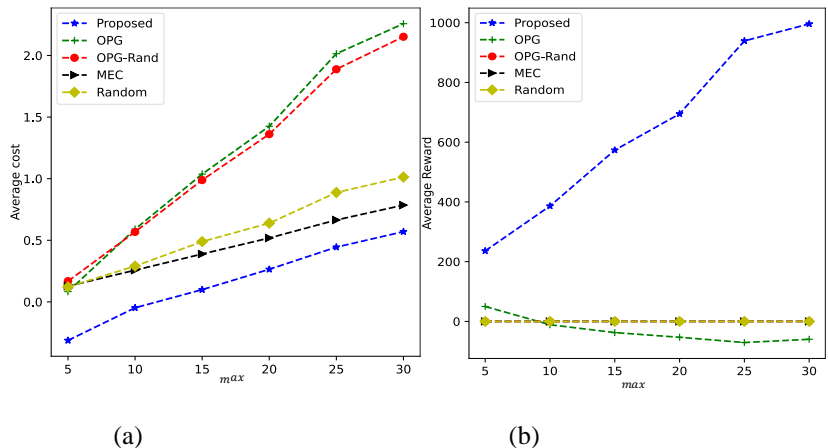

**Fig. 8.** Comparison of the influence of the maximum redundancy factor over each time slot: (a) average cost and (b) average reward.

### 2) *Influence of System Parameters*

We analyzed the influence of the COIN parameters on the computational cost from four aspects, including the redundancy factor, number of users, and number of subtasks. Fig. 8 illustrates the average cost and reward across time slots for the proposed and baseline schemes with varying $r_{max}$. While the overall cost increases by approximately 82% when $r_{max}$ goes from 5 to 30, the proposed method consistently reduces costs by an average of 59% compared to MEC, the most cost-effective baseline. In contrast, OPG and OPG-Rand experienced significant cost increases of 69% and 75% compared to the proposed approach. Additionally, the proposed system yields a 76% increase in reward as $r_{max}$ increases from 5 to 30. In contrast, OPG sees a substantial reward decrease of over 99% with higher $r_{max}$ values. These results emphasize the need for optimal redundancy factor estimation in BC-based offloading projects to enhance cost-effectiveness and reward compared to the current state-of-the-art OPG solution.

Furthermore, we evaluated the proposed model with varying users, and the results are as follows. In Fig. 9(a), our proposed method consistently proves its cost-effectiveness, reducing costs by 47% compared to the state-of-the-art OPG method as the user count increases. The OPG-Rand and random methods follow this cost reduction, while MEC's cost remains steady with increasing users. In terms of reward (Fig. 9(b)), the proposed agent achieves a remarkable 64% improvement over the OPG method, particularly with more than 20 users. This suggests that our approach enhances the OPG algorithm, making it more efficient in determining optimal offloading destinations while reducing costs and maximizing rewards.

Finally, we analyzed load distribution with varying numbers of subtasks. In Fig. 10(a), ten users with four subtasks distributed 75% to the EIN and the rest to the FIN. The OPG and OPG-Rand methods had similar distributions. With the EIN representing the MEC server, MEC showed a comparable pattern. The random method uniformly distributed tasks. Our approach, trained for 1000 epochs, reduced costs by 91%. In Fig. 10(b), ten users with seven subtasks evenly divided tasks between the EIN and FIN, resulting in a 63% cost reduction compared to random, with MEC being costlier. For ten subtasks with ten users, all methods except MEC performed poorly.

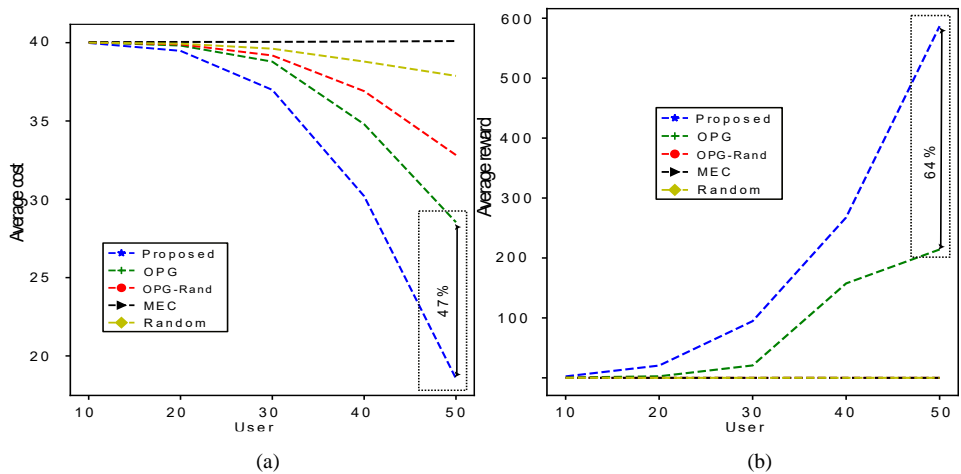

**Fig. 9.** Performance evaluation over each time slot against different numbers of users: (a) average cost and (b) average reward.

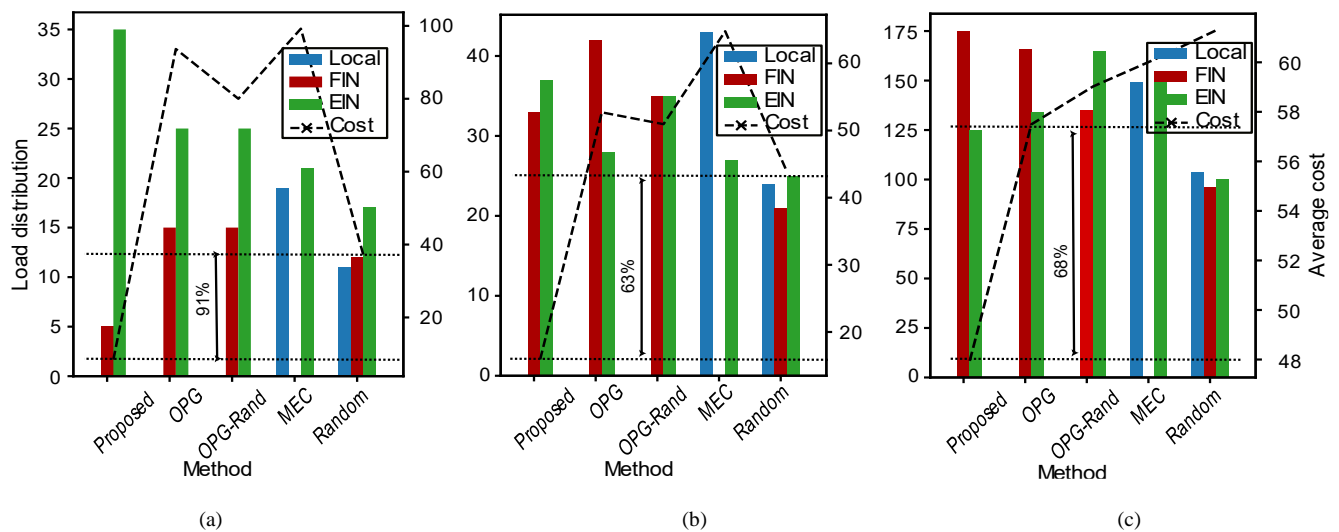

**Fig. 10.** Comparisons of the optimal partial computation offloading distribution under various numbers of subtasks: (a) ten users with four subtasks, (b) ten users with seven subtasks, and (c) 30 users with ten subtasks.

To achieve cost-effectiveness with ten subtasks, our approach required 30 users, as shown in Fig. 10(c). It reduced cost overhead by 68% compared to OPG, the second-best method. OPG-Rand, MEC, and random methods incurred increasing cost overhead. Except for MEC and random, all other algorithms executed no tasks locally, making OPG-based offloading suitable for metaverse applications with limited UE battery life.

### C. Discussion

In this research, as highlighted in section I, we proposed a redundancy-aware BC-based offloading approach to address scalability issues associated with BC-based partial offloading. This approach determines optimal redundancy factor updates to assist users in selecting the optimal offloading destination. We focused on divisible XR tasks, which are prevalent in metaverse applications. We formulated the offloading problem to offer an optimal offloading policy that considers dynamic user demand and computing resource availability.

Additionally, we explored the potential of COIN to address the computing and memory limitations faced by MEC [2, 54]. Moreover, existing studies have demonstrated the effectiveness of OPG solutions for task offloading [51], but they often focus on MEC without considering the BC scalability issues [15-20]. Our work bridges this gap by addressing BC data processing tasks and BC mining problems through partial offloading in COIN. We categorized computing into FIN and EIN to optimize latency and energy, considering dynamic user demand via the BC. Our investigation examined the impact of OPG-based PCO and BRF optimization using RL in COIN on delay and energy, effectively addressing BC scalability issues (see Section IV.).

This study offers a unique perspective on BC-based offloading, emphasizing the optimal redundancy factor's impact on communication and computing networks. We introduced the concept of a 'redundancy request' and employed DDQN to predict the optimal redundancy factor, helping users make tradeoffs between offloading cost, BC decentralization, and incentives, as demonstrated in Section IV-B.

### D. Limitations

Although the paper demonstrates the significance of redundancy-aware BC offloading in the metaverse, it is essential to note the scope limitations of this study. Firstly, the research does not involve the design or modification of consensus algorithms, which would be necessary for practical implementation, particularly regarding the assignment of optimal redundancy. Secondly, while our study considers energy and latency aspects, it does not encompass the energy consumption within the BC itself. Future research should incorporate an energy consumption model for the BC system. Lastly, the existing communication frameworks may have limitations for metaverse applications, and further investigation into task-oriented and sematic-aware communication is needed to optimize task allocation at the network level

## V. Conclusion

This paper investigated a joint BRF and PCO problem in a dynamic metaverse multiuser BC-enabled COIN consisting of EIN and FIN to minimize users' computation execution cost overhead while maximizing the incentive and satisfying the delay and BC offloading price constraints. Specifically, we solved the problem by splitting it into two stages. We first formulated the PCO problem as a multiuser PCO (OPG) game and proposed a decentralized algorithm to solve for its NE under any BC redundancy state. Then, we designed a DDQN-based BRF algorithm to solve for the optimal redundancy factors update for the BC. The BRF strategy is updated periodically based on the change in the user computational demand and dynamic network status to assist the PCO algorithm in making optimal offloading policy. The proposed scheme can capture metaverse divisible tasks, user communication and computing capacities, BC mining incentives, BC redundancy state, and inter-task redundancy correlation. The results demonstrated that the proposed method could effectively converge and predicts future optimal redundancy factor for future user offloading that minimizes the overall computation cost overhead while maximizing the BC mining incentives under delay and maximum incentive constraints.

## Appendix

### A. Proof of Lemma

We prove that problem $\mathcal{J}_p$ is NP-hard by demonstrating that the problem can be restricted to a maximum cardinality bin packing (MCBP) problem. In the MCBP problem, a given item $L$ with size $s_l$, $l \in \{1,2,..,L\}$, is assigned to $M$ bins of identity capacity $Q$ without exceeding the capacity constraint. The MCBP problem has been proven to be NP-hard [55]. To prove that the problem $\mathcal{J}_p$ contains the MCBP problem, we assume that the redundancy state of the remote resources and the user offloading demand is known in a particular time slot $t$ $(T = 1)$. The problem $\mathcal{J}_p$ is restricted as follows:

$$\widehat{\mathcal{J}_p} : \max_{\mathcal{S}_{k,t}} \sum_{k \in \mathcal{K}} \mathcal{C}_{k,t} \tag{38}$$

$$\text{s.t. 26c, 26f, and 26h.}$$

For problem $\widehat{\mathcal{J}_p}$, a user $k$ can benefit from remote execution

$(\mathcal{S}_{k,t} \in \mathcal{M})$ if and only if $\mathcal{C}_{k,t}^L \geq (1 - \beta_f^{(t)}) \mathcal{C}_{k,t}^O + \beta_f^{(t)} \mathcal{C}_{k,t}^R$; otherwise, it performs the computation locally $(\mathcal{S}_{k,t} = 0)$. Furthermore, the problem is restricted to remote execution $\mathcal{S}_{k,t} \in \mathcal{M}$ with the corresponding cost of -1 [2]: that is $(1 - \beta_f^{(t)}) \mathcal{C}_{k,t}^O + \beta_f^{(t)} \mathcal{C}_{k,t}^R = -1$. With $\mathcal{S}_{k,m}^{(t)} \in \{0,1\}$ where $\mathcal{S}_{k,m}^{(t)} = 1$ if and only if $\mathcal{S}_{k,t} = m$; otherwise, it is 0, the restricted problem $\widehat{\mathcal{J}_p}$ can be reformulated as

$$\widehat{\mathcal{J}_p} : \max_{\mathcal{S}_{k,t}} \sum_{k \in \mathcal{K}} \sum_{m \in \mathcal{M}} \mathcal{S}_{k,m}^{(t)} \tag{39}$$

$$\text{s.t. } \sum_{m \in \mathcal{M}} \mathcal{S}_{k,m}^{(t)} \leq 1 \tag{39a}$$

$$\sum_{k \in \mathcal{K}} \mathcal{S}_{k,m}^{(t)} \rho_k \eta_k \leq \psi \tag{39b}$$

$$\mathcal{S}_{k,m}^{(t)} \in \{0,1\} \tag{39c}$$

where the capacity $\psi$ is:
$$\psi = \mathcal{R}_{k,t} + \rho_k \eta_k$$
$$= \frac{\rho_k \eta_k}{\frac{M(I_f + V_f)(\delta_k^T + \delta_k^E \rho_k)}{2^{B\left(\delta_k^T\left(\tau_k^L - \tau_{k,FIN}^{exe}\right) + \delta_k^E \varepsilon_k^L + b_f^{(t)}\left(\delta_f^T \tau_{k,FIN}^{exe} - \delta_k^T \tau_{k,EIN}^{exe}\right)\right)} - 1} - \sigma^2}{+ \rho_k \eta_k.} \tag{40}$$

where (40) follows (28). Considering $\widehat{\mathcal{J}_p}$ as the MCBP, the items and bins correspond to the UE and channels. The size of item $l$ is $s_l = \rho_k \eta_k$. Therefore, if the problem $\widehat{\mathcal{J}_p}$ can be solved to determine the PCO strategy while stratifying the capacity constraint, a polynomial time algorithm can obtain the optimal solution to the MCBP problem. The original problem $\mathcal{J}_p$ can be regarded as an MCBP problem, which is NP-hard; thus, the problem is NP-hard.

### B. Proof of Remark 1

For a given PCO decision of all users except for user $k$, the modification of the strategy of user $k$ from $\mathcal{S}_{k,t}$ to $\mathcal{S}'_{k,t}$ culminates in a lower cost that leads to a decrease in the potential function. Consequently, the ordinal potential game $\mathcal{G}$ satisfies the following [56]:

$$\text{sgn}[\phi(\mathcal{S}_{k,t}, \mathcal{S}_{-k,t}) - \phi(\mathcal{S}'_{k,t}, \mathcal{S}_{-k,t})]$$
$$= \text{sgn}[f_t(\mathcal{S}_{k,t}, \mathcal{S}_{-k,t})$$
$$- f_t(\mathcal{S}'_{k,t}, \mathcal{S}_{-k,t})] \tag{41}$$

where sgn[.] denotes the sign function. The potential function $\phi(\mathcal{S}_{k,t}, \mathcal{S}_{-k,t})$ is derived as follows:

$$\phi(\alpha_{k,t}, \alpha_{-k,t})$$
$$= \frac{1}{2} \sum_k^K \sum_{n \neq k} \rho_k \eta_k \rho_n \eta_n \mathbb{1}(\mathcal{S}_{n,t} = \mathcal{S}_{k,t})(\mathcal{S}_{k,t} > 0)$$
$$+ \sum_{k=1}^K \rho_k \eta_k \lambda_k \mathbb{1}(\mathcal{S}_{n,t} = 0)$$
$$= \frac{1}{2} \sum_{n \neq k} \rho_k \eta_k \rho_n \eta_n \mathbb{1}(\mathcal{S}_{n,t} = \mathcal{S}_{k,t})(\mathcal{S}_{k,t} > 0)$$
$$+ \frac{1}{2} \sum_{l \neq k}^K \rho_l \eta_l \rho_n \eta_n \mathbb{1}(\mathcal{S}_{n,t} = \mathcal{S}_{l,t})(\mathcal{S}_{l,t} > 0)$$

$$+ \frac{1}{2}\sum_{l \neq k}^{K}\sum_{n \neq l, n \neq k} \rho_l \eta_l \rho_n \eta_n \mathbb{1}(\mathcal{S}_{n,t} = \mathcal{S}_{l,t})(\mathcal{S}_{l,t} > 0)$$
$$+ \rho_k \eta_k \lambda_k \mathbb{1}(\mathcal{S}_{k,t} = 0)$$
$$+ \sum_{k=1}^{K} \rho_l \eta_l \lambda_l \mathbb{1}(\mathcal{S}_{l,t} = 0)$$
$$= \rho_k \eta_k \sum_{n \neq k} \rho_n \eta_n \mathbb{1}(\mathcal{S}_{n,t} = \mathcal{S}_{k,t})(\mathcal{S}_{k,t} > 0)$$
$$+ \frac{1}{2}\sum_{l \neq k}^{K}\sum_{n \neq l, n \neq k} \rho_l \eta_l \rho_n \eta_n \mathbb{1}(\mathcal{S}_{n,t} = \mathcal{S}_{l,t})(\mathcal{S}_{l,t} > 0)$$
$$+ \rho_k \eta_k \lambda_k \mathbb{1}(\mathcal{S}_{k,t} = 0)$$
$$+ \sum_{k=1}^{K} \rho_l \eta_l \lambda_l \mathbb{1}(\mathcal{S}_{l,t} = 0). \tag{42}$$

To prove (41), we consider three cases: 1) $\mathcal{S}_{k,t} > 0, \mathcal{S}'_{k,t} > 0$; 2) $\mathcal{S}_{k,t} > 0, \mathcal{S}'_{k,t} = 0$; and 3) $\mathcal{S}_{k,t} = 0, \mathcal{S}'_{k,t} > 0$.

*Case 1):* $\mathcal{S}_{k,t} > 0, \mathcal{S}'_{k,t} > 0$.
Based on (42), we have
$$\phi(\mathcal{S}_{k,t}, \mathcal{S}_{-k,t}) - \phi(\mathcal{S}'_{k,t}, \mathcal{S}_{-k,t})$$
$$= \rho_k \eta_k \sum_{n \neq k} \rho_n \eta_n \mathbb{1}(\mathcal{S}_{n,t} = \mathcal{S}_{k,t})$$
$$- \rho_k \eta_k \sum_{n \neq k} \rho_n \eta_n \mathbb{1}(\mathcal{S}_{n,t} = \mathcal{S}'_{k,t})$$
$$= \rho_k \eta_k (\mathcal{R}_{k,t} - \mathcal{R}'_{k,t}). \tag{43}$$

According to (25), we have
$$f_t(\mathcal{S}_{k,t}, \mathcal{S}_{-k,t}) - f_t(\mathcal{S}'_{k,t}, \mathcal{S}_{-k,t})$$
$$= \sum_f \mathbb{1}(\mathcal{S}_{k,t} = 0)(I_f + V_f)(\delta_k^T + \delta_k^E \rho_k)\left(\frac{1}{\omega_{k,t}} - \frac{1}{\omega'_{k,t}}\right). \tag{44}$$

Based on (1) and (28), we obtain
$$\text{sgn}\left(\frac{1}{\omega_{k,t}} - \frac{1}{\omega'_{k,t}}\right) = \text{sgn}(\mathcal{R}_{k,t} - \mathcal{R}'_{k,t}).$$

Therefore, (41) is established for Case 1.

*Case 2):* $\mathcal{S}_{k,t} > 0, \mathcal{S}'_{k,t} = 0$.
Based on (42), we have
$$\phi(\mathcal{S}_{k,t}, \mathcal{S}_{-k,t}) - \phi(\mathcal{S}'_{k,t}, \mathcal{S}_{-k,t})$$
$$= \rho_k \eta_k \left(\sum_{n \neq k} \rho_n \eta_n \mathbb{1}(\mathcal{S}_{n,t} = \mathcal{S}_{k,t}) - \lambda_k\right)$$
$$= \rho_k \eta_k (\mathcal{R}_{k,t-} - \lambda_k). \tag{45}$$

According to (25), we have
$$f_t(\mathcal{S}_{k,t}, \mathcal{S}_{-k,t}) - f_t(\mathcal{S}'_{k,t}, \mathcal{S}_{-k,t})$$
$$= \sum_f \mathbb{1}(\mathcal{S}_{k,t} = 0)\left(\frac{(I_f + V_f)(\delta_k^T + \delta_k^E \rho_k)}{\omega_{k,t}} - (\delta_k^T T_k^L + \delta_k^E \mathcal{E}_k^L)\right). \tag{46}$$

Based on (28) analysis, we obtain
$$\text{sgn}(\mathcal{R}_{k,t-} - \lambda_k) = \text{sgn}\left(\frac{(I_f + V_f)(\delta_k^T + \delta_k^E \rho_k)}{\omega_{k,t}}\right.$$
$$\left. - (\delta_k^T T_k^L + \delta_k^E \mathcal{E}_k^L)\right).$$

Hence, (41) is established for this case.

*Case 3):* $\mathcal{S}_{k,t} = 0, \mathcal{S}'_{k,t} > 0$.
This case is similar to Case 2. Therefore, (42) is established. Thus, considering the results from the three cases, we conclude that the multiuser TS game $\mathcal{G}$ is an ordinal potential game and can achieve an NE solution in finite iterations [56].

### C. Proof of Lemma 2

Following [2], we analyze the convergence behavior of Algorithm 1 as follows. We let $\lambda_{max} = \max_{k \in \mathcal{K}}\{\lambda_k\}$, $\lambda_{min} = \min_{k \in \mathcal{K}}\{\lambda_k\}$, $\Omega_{max} = \max_{k \in \mathcal{K}}\{\rho_k \eta_k\}$ and $\Omega_{min} = \min_{k \in \mathcal{K}}\{\rho_k \eta_k\}$. Recalling the potential function in (30), we have

$$\phi(\mathcal{S}_t) \overset{(\text{U})}{\Rightarrow} \frac{1}{2}\sum_{k}^{K}\sum_{n \neq k} \rho_k \eta_k \rho_n \eta_n \mathbb{1}(\mathcal{S}_{n,t} = \mathcal{S}_{k,t})(\mathcal{S}_{k,t}$$
$$> 0) + \sum_{k=1}^{K} \rho_k \eta_k \lambda_k \mathbb{1}(\mathcal{S}_{k,t} = 0) \tag{47}$$
$$\leq \frac{1}{2}\sum_{k}^{K}\sum_{n \neq k} \Omega_{max}^2 \mathbb{1}(\mathcal{S}_{n,t} = \mathcal{S}_{k,t})(\mathcal{S}_{k,t} > 0)$$
$$+ \sum_{k=1}^{K} \Omega_{max} \lambda_{max} \mathbb{1}(\mathcal{S}_{k,t} = 0)$$
$$\leq \frac{1}{2} K^2 \Omega_{max}^2 + K \Omega_{max} \lambda_{max}$$

where (U) follows from (30). At the initialization stage, the user decisions are set to 0, which implies that the value of $\phi(0)$ is $\sum_{k=1}^{K} \rho_k \eta_k \lambda_k \geq K \Omega_{max} \lambda_{max}$. Hence, $\phi(\mathcal{S}_t) = \leq \frac{1}{2} K^2 \Omega_{max}^2 + K(\Omega_{max} \lambda_{max} - \Omega_{min} \lambda_{min})$. To lower the computing cost and, subsequently, the potential value, a user $k$ updates the decision $\mathcal{S}_{k,t}$ to a better $\mathcal{S}'_{k,t}$ in one iteration, that is, $\phi(\mathcal{S}_{k,t}, \mathcal{S}_{-k,t}) - \phi(\mathcal{S}'_{k,t}, \mathcal{S}_{-k,t}) > 0$. The decrement of $\phi(\mathcal{S}_t)$ in each iteration is analyzed in three cases: 1) $\mathcal{S}_{k,t} > 0, \mathcal{S}'_{k,t} > 0$; 2) $\mathcal{S}_{k,t} > 0, \mathcal{S}'_{k,t} = 0$; and 3) $\mathcal{S}_{k,t} = 0, \mathcal{S}'_{k,t} > 0$.

For Case 1) $\mathcal{S}_{k,t} > 0, \mathcal{S}'_{k,t} > 0$, we have
$$\phi(\mathcal{S}_{k,t}, \mathcal{S}_{-k,t}) - \phi(\mathcal{S}'_{k,t}, \mathcal{S}_{-k,t})$$
$$\overset{(\text{U})}{\Rightarrow} \rho_k \eta_k \sum_{n \neq k} \rho_n \eta_n \left(\mathbb{1}(\mathcal{S}_{n,t} = \mathcal{S}_{k,t}) - \mathbb{1}(\mathcal{S}_{n,t} = \mathcal{S}'_{k,t})\right) > 0 \tag{48}$$

where (U) follows from (43). For integer values of the indicator function $\mathbb{1}(.)$, we obtain
$$\sum_{n \neq k} \rho_n \eta_n \left(\mathbb{1}(\mathcal{S}_{n,t} = \mathcal{S}_{k,t}) - \mathbb{1}(\mathcal{S}_{n,t} = \mathcal{S}'_{k,t})\right) \geq \Omega_{min}. \tag{49}$$

Hence, $\phi(\mathcal{S}_{k,t}, \mathcal{S}_{-k,t}) - \phi(\mathcal{S}'_{k,t}, \mathcal{S}_{-k,t}) \geq \Omega_{min}^2$.

For Case 2) $\mathcal{S}_{k,t} > 0, \mathcal{S}'_{k,t} = 0$, we have
$$\phi(\mathcal{S}_{k,t}, \mathcal{S}_{-k,t}) - \phi(\mathcal{S}'_{k,t}, \mathcal{S}_{-k,t})$$
$$\overset{(\text{U})}{\Rightarrow} \rho_k \eta_k \left(\sum_{n \neq k} \rho_n \eta_n \mathbb{1}(\mathcal{S}_{n,t} = \mathcal{S}_{k,t}) - \lambda_k\right) \tag{50}$$

where (U) follows from (45). For a positive number $\pi = \sum_{n \neq k} \rho_n \eta_n \mathbb{1}(\mathcal{S}_{n,t} = \mathcal{S}_{k,t}) - \lambda_k$. Hence, $\phi(\mathcal{S}_{k,t}, \mathcal{S}_{-k,t}) - \phi(\mathcal{S}'_{k,t}, \mathcal{S}_{-k,t}) = \pi \rho_k \eta_k \geq \pi \Omega_{min}$.

For Case 3), $\mathcal{S}_{k,t} = 0, \mathcal{S}'_{k,t} > 0$ is similar to Case 2, where a positive number $\pi$ exists such that $\phi(\mathcal{S}_{k,t}, \mathcal{S}_{-k,t}) -$ $\phi(S'_{k,t}, S_{-k,t}) = \pi\rho_k\eta_k \geq \pi\Omega_{min}$. The potential function for the above three cases decreases by at least $\pi\Omega_{min}$ in each iteration. Consequently, the algorithm converges $\frac{\frac{1}{2}K^2\Omega_{max}^2 + K(\Omega_{max}\lambda_{max} - \Omega_{min}\lambda_{min})}{\pi\Omega_{min}}$ within the iteration and obtains an NE solution for the task offloading problem (TSP).

## ACKNOWLEDGMENT

The authors would like to express their gratitude to Sélinde Van Engelenburg for valuable feedback and contributions during the conceptualization phase.

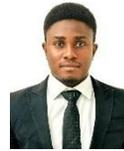

**Ibrahim Aliyu** (Member, IEEE) received a PhD in computer science and engineering from Chonnam National University, Gwangju, South Korea, in 2022. He also holds BEng (2014) and MEng (2018) degrees in computer engineering from the Federal University of Technology, Minna, Nigeria. He is currently a postdoctoral researcher with the Hyper Intelligence Media Network Platform Lab, Department of ICT Convergence System Engineering, Chonnam National University. His research focuses on distributed computing for massive metaverse deployment.

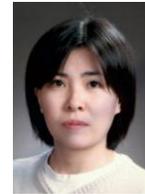

**Cho-Rong Yu** works in intelligence knowledge content research section at the ETRI. She received a bachelor's degree and master's degree in computer engineering from Chungnam National University in Daejeon, Korea. She is interested in virtual environment for training and new interfaces for AR/VR/XR environment including metaverse

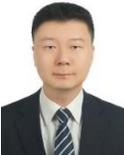

**Tai-Won Um** received a BS degree in electronic and electrical engineering from Hongik University, Seoul, South Korea, in 1999 and the MS and PhD degrees from the Korea Advanced Institute of Science and Technology (KAIST), Daejeon, South Korea, in 2000 and 2006, respectively. From 2006 to 2017, he was a principal researcher with the Electronics and Telecommunications Research Institute (ETRI), a leading government institute on information and communications technologies in South Korea. He is currently an associate professor at Chonnam National University, Gwangju, Korea. He has been actively participating in standardization meetings, including ITU-T SG20 (Internet of Things, smart cities and communities).

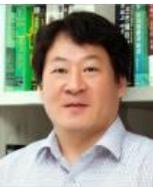

**Jinsul Kim** received a BS degree in computer science from the University of Utah, Salt Lake City, Utah, USA, in 1998 and MS (2004) and PhD (2008) degrees in digital media engineering from the Korea Advanced Institute of Science and Technology (KAIST), Daejeon, South Korea. Previously, he worked as a researcher at Broadcasting/Telecommunications Convergence Research Division, Electronics and Telecommunications Research Institute (ETRI), Daejeon, Korea, from 2004 to 2009 and was a professor at Korea Nazarene University, Cheonan, Korea, from 2009 to 2011. He is a professor at Chonnam National University, Gwangju, Korea. He is a member of the Korean national delegation for ITU-T SG13 international standardization. He is a co-research director of the AI Innovation Hub Research and Development Project hosted by Korea University and the director of the G5-AICT research center.